\documentclass{aa}
\usepackage{graphicx}
\usepackage{amsmath}
\usepackage{txfonts}
\usepackage{rotating}
\begin{document}
   \title{Searching for star-forming galaxies \\ in the Fornax and Hydra clusters 
          \thanks{Based on observations acquired at Gemini South (GS-2007B-Q-53 and GS-2008A-Q-49) and ESO VLT (382.A-0409A)}
         }

      \author{O. Vaduvescu\inst{1,2}
          \and
              C. Kehrig\inst{3}
          \and
              J. M. Vilchez\inst{4}
          \and
              E. Unda-Sanzana\inst{2}
              }

   \offprints{O. Vaduvescu}

   \institute{Isaac Newton Group of Telescopes, Apto. 321, E-38700 Santa Cruz de la Palma, Canary Islands, Spain\\
              \email{ovidiuv@ing.iac.es}
             \and
              Instituto de Astronom\'ia, Universidad Cat\'olica del Norte, Avenida Angamos 0610, Antofagasta, Chile\\
            \and
              Leibniz-Institut fur Astrophysik Postdam, innoFSPECPostdam, An der Sternwarte16, 14482 Postdam, Germany\\
            \and
              Instituto de Astrof\'isica de Andaluc\'ia (CSIC), Apto. 3004, 18080, Granada, Spain\\
             }

   \date{Astronomy and Astrophysics 533, A65 (2011)}

% \abstract{}{}{}{}{} 
% 5 {} token are mandatory
 
  \abstract
  { 
   The formation and evolution of dwarf galaxies is relatively difficult to understand because 
   of their faint emission in all regimes that require large aperture telescopes. 
  } 
  { We intend to study the evolution of star forming dwarf galaxies in clusters. We selected Fornax and 
   Hydra clusters to complement our previous study of Virgo. On the basis of available literature data, 
   we selected ten star-forming candidates in Fornax and another ten objects in Hydra. } 
  { We used Gemini South with GMOS to acquire $H\alpha$ images necessary to detect star-forming regions 
   in the two galaxy samples. We then performed long-slit spectroscopy for the brightest six candidates, 
   to derive their chemical properties. Finally, we employed the VLT with HAWK-I to observe all galaxies 
   in the $K^\prime$ band to derive their main physical properties. } 
  { We studied the morphology of our two samples, finding five objects in Fornax and six in Hydra 
   with structures consistent with those of star-forming dwarfs, i.e., dwarf irregulars (dIs) or blue 
   compact dwarfs (BCDs). 
   About four other objects are probably dwarf spirals, while three objects remained undetected in both 
   visible and near infrared. 
   On the basis of visible bright emission lines, we derived oxygen abundances for ten star-forming 
   candidates with values between 8.00 $\le$ 12+log(O/H) $\le$ 8.78. 
  } 
  { Most fundamental properties of star-forming galaxies in Fornax and Hydra appear similar to 
   corresponding properties of dIs and BCDs from Virgo and the Local Volume (LV). 
   The luminosity-metallicity and metallicity-gas fraction relations in the LV and Virgo appear to be followed 
   by Fornax and Hydra samples, suggesting that the chemical evolution of the two clusters seems consistent 
   with the predictions from the closed box model, although larger samples are needed to investigate the role of 
   possible environmental effects. 
   Star-forming dwarfs (dIs and BCDs) in different environments appear to follow different mass-metallicity relations, 
   with more metal-rich objects tending to occupy regions of higher galaxy overdensity in the Virgo and Hydra clusters. 
   Treated separatelly, dIs and BCDs also appear to define different mass-metallicity relations, with BCDs following a 
   steeper relation than dIs fitted alone. 
   Only two objects from Fornax and Hydra appear confined to the dwarf fundamental plane (FP) which does not 
   seem to hold for most Hydra objects, suggesting that the environment has some influence. 
   A concerted effort is necessary to acquire deep radio data for these nearby clusters in conjunction with NIR 
   imaging and spectroscopy. 
   } 


   \authorrunning{Vaduvescu et al.}
   \maketitle

%
%________________________________________________________________

\section{Introduction}

The concept of ``compact galaxies'' was introduced by Zwicky (\cite{zwi65,zwi70}) to describe 
any galaxy whose surface brightness is brighter than 20 mag arcsec$^{-2}$ in any wavelength 
range. The term ``blue'' later used refers to compact galaxies observed on blue and red 
plates (Zwicky \cite{zwi70}; Zwicky \& Zwicky \cite{zwi71}), thus blue compact galaxies 
must be blue in colour. 
Thuan \& Martin (\cite{thu81}) later introduced the term ''blue compact dwarf'' (BCD) 
to characterize three main properties of compact galaxies: their low luminostity ($M_B\geq-18$), 
small size (diameter $\leq \sim 5$ kpc), and strong narrow emission lines superposed on a nearly 
flat continuum. The star-forming rates of BCDs are very high (between 0.1 and 1 $M_\odot$ yr $^{-1}$) 
in comparison with those in dwarf irregular galaxies (dIs), while their metallicities are low 
(oxygen abundance between 1/50 and 1/2 $Z_\odot$) relative to with giant galaxies, which 
implies that they have young ages (Kunth \& Ostlin, \cite{kun00}). 

Despite much research in the past three decades, many aspects of the evolution of star 
forming galaxies remain unclear. Thuan (\cite{thu85}) suggested that BCDs may be bursting dIs, 
while Kunth \& Ostlin (\cite{kun00}) argued that the gas of dIs is insufficient to generate a 
starburst bright enough to transform a dwarf into a BCD. Richer \& McCall (\cite{ric95}) 
proposed that the descendants of BCDs may be dwarf spheroidal galaxies (dSphs) rather than 
dIs, while James (\cite{jam94}) argue that dIs are fundamentally distinct from BCDs, with no 
simple evolutionary connection between the two classes. 

Star-forming dwarf galaxies appear to be quite abundant in the Universe. Of the 451 galaxies in the 
Local Volume (LV, $D<10$ Mpc), about 70\% (315 galaxies) have late Hubble types ($T=9$ or $10$), 
most being dI candidates (Karachentsev et al. \cite{kar04}). About 64 BCDs and BCD candidates were 
classified in the Virgo cluster (Bingelli, Sandage \& Tammann \cite{bin85}), while about 114 BCDs 
are listed as closer than 57 Mpc (Paz, Madore \& Pevunova \cite{paz03}). Unfortunatelly, to date there 
has been no similar census in other nearby clusters. We know that the environment of galaxies can influence 
many processes regulating their star formation activity, hence their photo-chemical evolution. 
Therefore, environmental parameters of star-forming galaxies should be taken into account when 
trying to  understand better the star formation process. The observation of star-forming dwarf 
galaxies populating some dense cluster neighborhood should shed light on the role played by the 
environment on their evolution.  

Using facilities in the North, we have studies star-forming dwarf galaxies located both in the 
LV and the Virgo cluster to scrutinize their physical and chemical properties and compare their 
evolution (Vilchez \& Iglesias-Paramo \cite{vil03}; Vaduvescu et al. \cite{vad05}; 
Vaduvescu, Richer \& McCall \cite{vad06}; Vaduvescu, McCall \& Richer \cite{vad07}). 
BCDs and dIs appear to be physically related, showing similar correlations between size, 
near-infrared (NIR) luminosity, mass and internal energy, and populating the same ``fundamental plane'' 
(Vaduvescu \& McCall \cite{vad08}) which also appears to be shared by dwarf ellipticals (dEs). 
The chemical properties of BCDs and dIs also confirm a possible link between the two classes via 
the metallicity-luminosity relation ($L-Z$), which appears to be tighter in NIR than in visible, 
although BCDs appear to share a $L-Z$ relation steeper than dIs. The position of a dwarf on the $L-Z$ 
diagram may indicate that there is a relation between the depth of the potential well and the 
dwarf's ability to retain gas during the course of evolution. Nevertheless, to examine the 
prevalence of past gas flows, it is necessary to go beyond the $L-Z$ relation and study the 
relationships between metallicity, gas fraction, and mass. 

The most popular framework of galaxy evolution is the ``closed box model'' (Searle \& Sargent 
\cite{sea72}; Pagel \cite{pag97}). According to this model, a galaxy consists initially of 
gas with no stars and no metals. The stellar initial mass function (IMF) is assumed to be
constant with time. Stars that end their life are assumed to enrich immediately the interstellar 
gas with metals, and throughout its entire life the galaxy evolves as a closed system with no 
mass inflow or outflow. 
In a simple closed box model, relations between metallicity, gas fraction, and mass have 
been proposed to exist (Pagel \cite{pag97}; Lee et al. \cite{lee03}; Vaduvescu, McCall \& 
Richer \cite{vad07}) but to decouple the closed box model from the environmental 
influences, we need to study star-forming dwarfs in a broad range of clusters. 

The metallicity of BCDs is commonly defined as the oxygen abundance of their HII regions 
detected using $H\alpha$ observations. Abundances can be measured precisely using the direct 
method based on the faint [OIII]$\lambda$4363 line (Osterbrock \cite{ost89}) or less precisely 
using a few bright line methods (Pagel \cite{pag97}; McGaugh \cite{mg94}; Pilyugin \cite{pil00}). 
The direct method is more difficult to apply to distant dwarfs, although accurate results in 
Virgo have been obtained using the 4m WHT telescope (Vilchez \& Iglesias-Paramo \cite{vil03}) 
and the 8m Gemini North telescope (Vaduvescu, Richer \& McCall \cite{vad06}) which have  
encouraged us to measure abundances of dwarfs in other nearby clusters using 8m telescopes. 
Studying galaxies in clusters, we have been able to reduce the scatter in distance-dependent 
parameters such as the absolute magnitude and mass. 
In contrast, if we had selected candidates in low density intergalactic regions, one could 
of course avoid the perturbation effects of the dense environment, being able to consider the 
simple closed box model regime. 

This is the beginning of a long-term project to study star-forming galaxies in clusters 
and compare their properties and evolution with that of isolated objects in the LV. 
In the present paper, we scrutinize two small samples of star-forming candidate galaxies, specifically 
dwarfs (dIs and BCDs) located in two southern clusters, namely Fornax and Hydra. On the basis of the available 
literature, we selected and observed in each cluster ten star-forming candidates using Gemini with GMOS 
for $H\alpha$ pre-imaging and long-slit spectroscopy and VLT with HAWK-I for NIR imaging. 
In Section~\ref{observations}, we present the observations and Section~\ref{results} includes the results. 
In Section~\ref{discussion}, we discuss morphological aspects, chemical abundance analysis, and the 
fundamental relations of the two cluster samples. Finally, in Section~\ref{conclusions} we draw the 
conclusions.

%__________________________________________________________________

\section{Observations}
\label{observations}

This paper represents our first attempt to compare galaxy formation and evolution in 
nearby clusters. Vaduvescu, McCall \& Richer (\cite{vad07}) used spectroscopy acquired with Gemini North 
and other data taken with 4m WHT telescope (Vilchez and Iglesias-Paramo, \cite{vil03}) to study a 
sample of star-forming dwarfs in Virgo and compare its evolution with a field sample. To address 
possible environmental effects, star-forming dwarfs located in other clusters should be studied 
further, and the most accessible clusters remain the closest ones ($d<100$ Mpc). Nevertheless, 
emission from dwarfs located at these distances is extremely difficult to detect, hence 
8-10m class telescopes are necessary for this work. 

\subsection{Cluster selection and membership}

Fornax is the second closest cluster after Virgo and is located at $20.0 \pm 0.2 \pm 1.4$ Mpc 
($DM=31.51$) according to a measurement by the surface brightness fluctuations method using HST/ACS 
data (Blakeslee et al. \cite{bla09}) which closely matches a previous Cepheid value derived by 
Madore et al. (\cite{mad98}). Although smaller and poorer in terms of members than the Virgo 
cluster, Fornax is more compact than Virgo and is expected to include about 340 likely members in its central 
$\sim40$ square degree region (Ferguson \cite{fer89}). 

Hydra cluster is located at $46 \pm 5$ Mpc ($DM=33.31$) based on the infrared surface 
brightness fluctuation method (Jensen, Tonry \& Luppino \cite{jen99}), thus appears to be 
the fourth closest cluster after Virgo, Fornax and Antlia. Easily observable from 
the south hemisphere, Fornax and Hydra are natural choices for the search and study of star-forming 
dwarf galaxies in different environments. For each cluster, we selected our targets 
based on existing papers and data available from the literature. 

A complete $H\alpha$ coverage of the two clusters should cover some 40 square degrees, 
and is virtually impossible to conduct using regular application time. For the sample selection, 
we took into account the scarce X-ray mapping available in the literature, to help us to 
avoid high density environments. Unfortunately, these maps are small and cover only the central 
regions of the clusters. 
For Fornax, we examined the Chandra X-ray image of the central field covering the 
$47\arcmin$ square field close to the giant elliptical NGC 1399 (Chandra website
\footnote{http://chandra.harvard.edu/photo/2004/fornax/more.html}
concluding that all our ten sample objects are located outside of this field. 
For Hydra, we used the ROSAT X-ray map containing the central $18\arcmin$ field of the cluster 
close to the giant elliptical NGC 3311 (NASA, \cite{nas03}) and the $20\arcmin$ X-ray 
map from Fitchett \& Merritt (\cite{fit88}), neither of which include any of our sample objects. 
Despite the uncertain mapping of our sample with respect with the X-ray cluster density, 
our datasets represent a step forward in studying star-forming galaxies in two neighbouring  
clusters, being complementary to some previous work (Ferguson \cite{fer89}, Drinkwater et 
al \cite{dri01a,dri01b}, Duc et al. \cite{duc99,duc01}). 

\subsubsection{Fornax cluster}

The Fornax Cluster Catalog (FCC - Ferguson \cite{fer89}) was constructed based on the visual 
inspection of the photographic plates taken with the 2.5m du Pont telescope at Las Campanas 
Observatory. The catalog lists 340 likely members and 2338 likely background galaxies. 
Because very few radial velocities were available, the criteria used to distinguish cluster 
members from background galaxies were based mostly on visual morphological analyse. The 
author identifies in the catalog a total of 35 BCD candidates, but notes that there are 
very few actual members of the cluster, classifying only five BCDs as definite or likely 
members while another 30 appear as dI candidates (classes ImIV or ImV). 

Drinkwater et al. (\cite{dri01a}) acquired multi-fiber spectroscopy with the 1.2m UK Schmidt 
telescope to study the evolution and star formation of dwarf galaxies in the central 6 degree 
field of the Fornax cluster. Their work focused on compact galaxies in the search for new 
BCDs that were not included in the FCC. They measured redshifts for 516 galaxies of which 
108 were found to be members of the cluster. The spectroscopy of 19 FCC candidate BCDs confirm the 
morphological FCC findings that most candidates are background galaxies, with only three cluster 
members being detected to have $H\alpha$ emission. Only a few (12) objects were measured to have 
different membership classifications from those of the FCC. 

Schroder, Drinkwater \& Richter (\cite{sch01}) presented a new set of HI observations of 
member galaxies of the Fornax cluster using the Parkes radio telescope. This paper provides 
the HI fluxes, gas masses, and $W_{20}$ velocity widths of 66 galaxies, including 5 BCD/Im 
candidates some having $W_{20}$ measurements. Taking into account these literature data 
and the morphological appearance of all candidates, we selected ten objects to be observed 
in the Fornax cluster, which are included in the first part of Table~\ref{table1}. 

\subsubsection{Hydra Cluster}

Studies of dwarfs in Hydra cluster are scarcer than for Fornax, most probably owing to 
Hydra's distance of more than twice that of Fornax. Duc et al. (\cite{duc99}) presented 
HI, optical, and near-infrared photometric data of a sample of HI-selected dwarf galaxy 
candidates in Hydra. Their optical imaging was taken in the $B$ and $I$ bands with the 1.5m 
Danish telescope in La Silla, the NIR imaging in $K$ band with the ESO/MPG 2.2m telescope, 
and the radio data was acquired at high resolution with the Nancay decimetric radiotelescope 
to follow up a former VLA survey. 

A second paper of the same authors (Duc et al. \cite{duc01}) presents a spectroscopic 
follow-up of a sample of 15 HI-selected dwarf galaxies in Hydra cluster. Long-slit 
spectra were obtained along an axis including the nucleus and the brightest HII region 
with the ESO 3.6m telescope. Only in one galaxy could the faint [OIII]$\lambda$4363 line 
be detected to derive oxygen abundances via the direct method, while for ten other dwarfs 
the abundance 
could be calculated via other bright line methods and for four others no line detection was 
possible. On the basis of the analysis of the surface brigtness profiles, existing spectroscopy, 
radio data ($W_{20}$ line) and their close morphological inspection, we selected 10 objects 
to be observed in Hydra. They are included in the second part of Table~\ref{table1}. 

\subsection{Gemini $H\alpha$ Imaging and Spectroscopy}

Two proposals were observed by Gemini South in queue mode in 2007B for Fornax (Run ID: 
GS-2007B-Q-53) and in 2008A for Hydra cluster (Run ID: GS-2008A-Q-49). During both runs, 
we used the GMOS South camera (GMOS-S) to acquire $H\alpha$ imaging of ten star-forming 
candidates in Fornax and ten candidates in the Hydra cluster. The GMOS-S camera consists 
of three CCDs $2048 \times 4608$ pixels each with pixel size 
$13.5~\mu$m, resulting in a $6144 \times 4608$ pixel mosaic covering 
$5.5\arcmin \times 5.5\arcmin$ on-sky field. 
All observations were taken in band 3 weather conditions, using $4 \times 4$ binning mode 
(pixel scale of $0 \farcs 292$) that yielded a total field of view of $10\farcm0 \times 7\farcm5$. 
For all targets, we used the Ha\_G0336 (referred to as $H\alpha$ from now on) and r\_G0326 
(referred to as $R$ from now on) Gemini filters. Gemini imaging is presented in 
Figure~\ref{fig1} and the observing log is given in Table~\ref{table1}. 

We processed the pre-imaging frames (only the central CCD containing all galaxies) 
using the appropriate bias and flat field (in $H\alpha$ and $R$), then combined the 
individual frames to produce the final reduced images. The ``net'' $H\alpha$ images were 
reduced via the usual procedure measuring about 20-30 stars, whose measured magnitudes we 
scaled in order to subtract $R$ images from the original $H\alpha$ images. 
The runs were observed in non-photometric band 3 
conditions, so for the $R$ images we derived zero points using $10-20$ stars in the observed 
fields with available $R$ magnitudes from the USNO-B1 catalogue (Monet et al. \cite{mon03}). 
The USNO-B1 magnitudes are in the photographic system and given at two different epochs, 
hence we averaged the two values. We acknowledge the possible need for a small shift 
between photographic and Gemini $R$ system, but we report that we found our total $R$ 
magnitudes to be consistent, within the errors, with those available in the literature. 

The spectroscopic raw frames were processed using the Gemini GMOS routines within 
IRAF\footnote{IRAF is distributed by the National Optical Astronomical Observatories, which 
are operated by the Association of Universities for Research in Astronomy, Inc., under 
cooperative agreement with the National Science Foundation.}. Biases and flat-fields were 
combined with the tasks ``gbias'' and ``gsflat'', respectively. 
The following standard stars were observed: LTT 1788, LTT 1020, LTT 4364, and LTT 1377 for the 
Fornax run (one star observed every science night), while LTT 3218 and LTT 7378 were observed 
for the Hydra run (each in one night). 
Science targets and standard stars were reduced with task ``gsreduce'', which applies the overscan 
correction, subtracts off the bias, mosaics the three detectors of GMOS interpolating across the chip 
gaps for science data, and applies flat-field correction. The wavelength calibration was established 
from the CuAr arcs with the task ``gswavelength'' giving residuals $<$ 0.1~\AA. The science 
spectra were wavelength calibrated by applying the task ``gstransform''. We use the 
task ``gsextract'' to extract the one dimensional spectra performing sky subtraction. The 
sensitivity function was created by using the task ``gsstandard'', and finally we flux 
calibrated the science images making use of the task ``gscalibrate''. The statistical uncertainty 
in the instrument response fitting across the spectrum from blue to red sides is $\sim 2 \%$.

\subsection{VLT NIR Imaging}

$K^\prime$ imaging was observed in service mode at VLT in 2008B for Fornax and Hydra 
clusters (Run ID: 382.A-0409). We used the HAWK-I camera, which consists of 4 HAWAII 2RG 
detectors with a pixel scale of $0.106\arcsec$ resulting in a $4096\times4096$ pixel mosaic 
covering $7.5\arcmin\times7.5\arcmin$ field of view. For all targets, we used the $K^\prime$ 
filter. 

The image reduction was done using REDNIR, an IRAF script that subtracts each sky image compiled 
from two neighbouring galaxy images, allowing a close temporary and spatially mapping of the 
rapid variation in the sky level and pattern (Vaduvescu and McCall \cite{vad04}). The reduced 
VLT imaging is included in Figure~\ref{fig2} and the observing log is given in Table~\ref{table1}. 
To reduce our $K^\prime$ photometry, we calculated zero-points based on the available 2MASS stars 
visible in the observed fields. The number of available 2MASS stars was about five in each 
Fornax field and around ten in the denser Hydra fields. In a few cases, only two or three 
2MASS stars were available, but the zeropoints are consistent with the remainder, with both 
sets of fields having a standard deviation around 0.1 mag.

%______________________________________________________________

\section{Results}
\label{results}

\subsection{Imaging}

We present next the imaging observations acquired at Gemini (in the visible) and VLT (in the NIR). 

\subsubsection{R and $H\alpha$ Imaging}

\begin{figure}[p]
\centering
\includegraphics[angle=0,width=14cm]{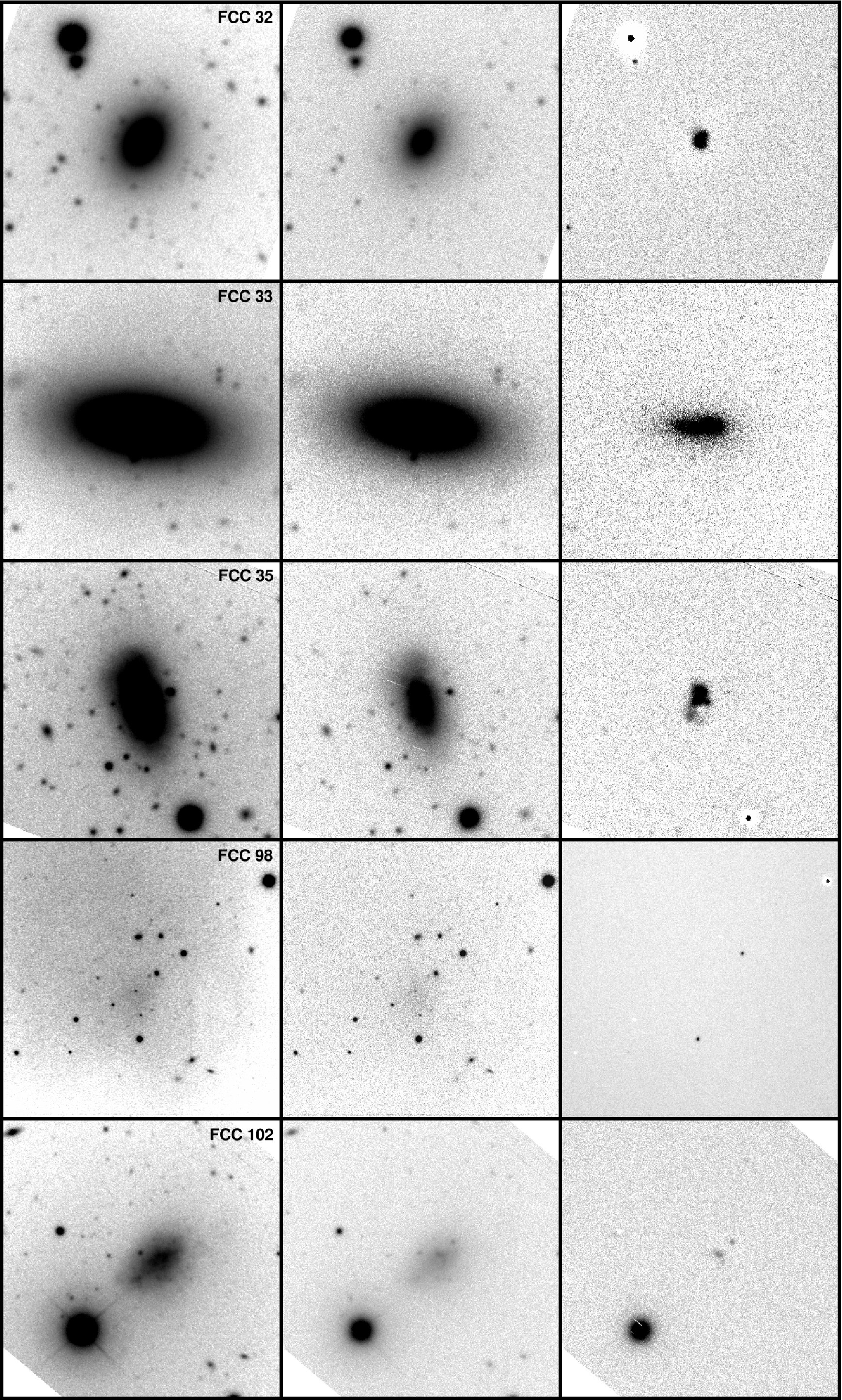}
\begin{center}
\caption{$H\alpha$ (Ha\_G0336 - left column), continuum (r\_G0326 - mid column) and pure $H\alpha$ 
images (right column) of the Fornax and Hydra star-forming candidates observed at Gemini South. 
Field of view is $2\arcmin \times 2\arcmin$, normal orientation (north above, east at left). } 
\label{fig1}
\end{center}
\end{figure}

For the $H\alpha$ pre-imaging phase, we observed the whole sample with Gemini, namely 10 galaxies 
in Fornax and 10 in Hydra. Images in $R$, $H\alpha$, and reduced $H\alpha-R$ for both clusters are 
presented in Figure~\ref{fig1}. 

Only six galaxies of Fornax have detectable net $H\alpha$ emission in the right panel of Figure~\ref{fig1}. 
In FCC~32, FCC~33, and FCC~35, some bright compact $H\alpha$ emission is clearly visible at their 
centres, with FCC~33 having the brightest core elongated in the east-west direction and FCC~35 being
surrounded by a diffuse envelope that includes three small knots to the south and west of the core. 
FCC~102 and FCC~120 clearly display two separated knots, more compact and closed in the first case and 
more diffuse and separated in the second; a rather similar distribution can be seen for the 
three central knots of FCCB~905. The galaxy FCC~128 seems to contain a very faint and diffuse  
component, which is however unmeasureable about the noise level. The relevant physical properties 
of the Fornax $H\alpha$ selected galaxies are presented in Table~\ref{table2}. 
 
Although all Hydra galaxies have different morphologies and rich details in the $R$ and $H\alpha$ 
frames presented in Figure~\ref{fig1} (most of them diffuse in $H\alpha$), they show no 
clear net measurable $H\alpha$ emission at the limit of detection of this work; this is possibly 
a consequence of either the poor weather observing conditions, an imperfect continuum removal, or both. 
Nonetheless, we should bear in mind that our Hydra targets are three times more distant than Fornax. 

On the basis of the $R$ images observed in relatively good conditions with Gemini, we studied the 
$R$ surface photometry and colors of all 20 objects. By using both Gemini $R$ and the VLT $K^\prime$ 
surface photometry taken with similar aperture and resolution instruments and assuming $K^\prime$ to be 
more fundamental than $R$ in tracing mass, we derived the $R$ profiles based on the fitted parameters 
in $K^\prime$. Using both $R$ and $K^\prime$ photometry, we compiled the $R-K^\prime$ color profiles 
of all detected galaxies. 

The zero points in $R$ were derived using stars in the field with magnitudes taken from the 
USNO-B1 catalog, so these should be regarded with caution, although we note that  
the color profiles are unaffected by the $R$ zero point. The surface brightness profiles in $R$ 
and $R-K^\prime$ color profiles are included in the right panel of Figure~\ref{fig2}. 
The derived total $R$ apparent magnitudes and the physical properties of Hydra galaxies are presented 
in Table~\ref{table2}. 

\subsubsection{$K^\prime$ imaging}

For the NIR imaging, we observed the whole sample with the VLT, namely 10 galaxies in Fornax and 
10 in Hydra. To perform the surface photometry, we employed the STSDAS package (ELLIPSE task) in IRAF 
using an iterative approach that converged for fixed parameters (centre, ellipticity, and position 
angle), which are common for the two ($K^\prime$ and $R$) bands. We initially derived fitting parameters 
using $K^\prime$ images that we later compared to the parameters obtained from our $R$ fits. 
Within our estimated errors (about 0.5 arcsec in position for centres, 0.1 in ellipticity, and 
5 degrees in position angle), the fits in $K^\prime$ and $R$ bands agree very well. 
The sky contibution in the visible is much fainter than in the NIR, so profiles in 
Figure~\ref{fig2} in $R$ are deeper than the corresponding ones in $K^\prime$ (the profiles in 
$R$ extend outside our plotted common region), while the associated error bars are much 
smaller in $R$ (smaller than the symbols used). 

To fit the $K^\prime$ surface brightness profiles of star-forming dwarf galaxies, we used the 
same approach as in our previous work, by modeling profiles of dIs with a hyperbolic secant (``sech'') 
function (Vaduvescu et al. \cite{vad05}) and those of BCDs with a sech function plus a Gaussian 
(Vaduvescu, Richer \& MacCall \cite{vad06}). In this sense, we used the NFIT1D function of FITTING package 
under IRAF. In Figure~\ref{fig2}, we plot with a solid continuous line the sech fit 
(for dIs and BCDs), with a dashed line the Gaussian (for BCD-like objects only), and with a thin line the 
total that closely follows the profiles of BCDs (see for example the plots for FCCB~905 or H1031-2632). 

In Table~\ref{table2}, we include our derived physical parameters for the galaxies detected in $K^\prime$ 
in Fornax and Hydra. Ellipticity $e$ and position angle $PA$ (measured positively counter-clockwise) 
are given in the first two columns. Total magnitudes $m_{TK}$ were measured by integrating growing 
ellipse apertures, while sech magnitudes $m_{SK}$ were derived by modeling the profiles with the 
sech (where $\mu_{0K}$ represent its central surface brightness expressed in mag/arcsec$^2$ and 
$r_{0K}$ is the scale radius expressed in arcsec) to which we eventually add a Gaussian function 
only for BCD-like objects to model a central outburst. The calculated semimajor radius $r_{K22}$ (in 
arcsec) at $m_K=22$ mag/arcsec$^2$ are given next. 
We provide in column 9 the total magnitudes in $R$ band, which are 
measured in a similar way integrating growing ellipse apertures. Absolute magnitudes in $K^\prime$, 
M$_{SK}$ (column 10), were calculated from sech $m_{SK}$ using the published distance modulus (DM) for the two clusters, 
namely 31.51 for Fornax (Blakeslee et al. \cite{bla09}) and 33.31 for Hydra (Jensen, Tonry \& Luppino \cite{jen99}). 
The $W_{20}$ HI velocity line widths, quoted in column 11 (in km/s), were taken from the HYPERLEDA 
database, and the original reference for these data is given in column 12. 
The logarithms of the galaxy stellar mass (column 13), gas mass (column 14), and baryonic mass (column 15) 
all in solar units (M$_{\odot}$) are presented, as well as the gas fraction $\mu$ (column 16), 
calculated following Vaduvescu, McCall \& Richer (\cite{vad07}). 

\subsection{Spectroscopy}

Taking into account our $H\alpha$ fluxes and the complementary information available in the 
literature (abundances and radio data) and on the basis of the time available to perform spectroscopy, 
we selected six star-forming dwarf candidates in Fornax and six similar objects in Hydra to be 
observed spectroscopically with Gemini. 
From these, we obtained emission line spectra with the necessary signal-to-noise ratio for all selected 
Hydra objects and only four observed Fornax galaxies. In the following section, we present our results 
obtained from spectroscopy. 
 
\subsubsection{Line fluxes}

Emission-line fluxes were measured in the one-dimensional spectra extracted for each object using 
the IRAF task \textsc{SPLOT}, which integrates the line flux over a locally fitted continuum. 
To minimize the errors 
associated with the relative (to H$\beta$) flux calibration over the full wavelength range 
(especially for faint lines), line ratios were referred to the flux of the nearest 
Balmer line, H$\beta$ or H$\alpha$. 

Figure~\ref{fig3} displays the calibrated spectra of the galaxies of Fornax
and Hydra clusters for which spectroscopic observations were sucessful. We provide some illustrative 
spectra for our sample which include FCC~35 (Fornax) exibiting a typical HII-region-like spectrum with 
a very faint continuum and strong emission lines, and H1031-2818 (Hydra) displaying absorption in the 
Balmer series and faint emission lines. 

We derived the logarithmic reddening constant, C(H$\beta$), from the ratios of the optical Balmer 
recombination lines, while simultaneously solving for the effects of underlying Balmer absorption 
using equivalent width, EW$_{abs}$. We assumed that the equivalent widths of the main Balmer absorption 
lines were approximately the same, as expected for young ionizing clusters (Kobulnicky \& Skillman \cite{ks96}; 
Kobulnicky, Kennicutt \& Pizagno \cite{k99}). The correction for underlying absorption ranged 
from equivalent widths equal to zero up to values around 8 \AA . The values derived for C(H$\beta$) 
vary from nearly zero to 0.9. 
The line-flux errors were calculated using the expression (e.g. Kehrig et al, \cite{keh08})

\begin{equation} 
\sigma_{line}=\sigma_{cont}N^{1/2}\left(1 + \frac{\rm EW}{N\Delta\lambda}\right)^{1/2} ,
\end{equation}
\noindent
where $\sigma_{cont}$ is the standard deviation in the continuum near the emission line, $N$ is 
the width of the region used to measure the line in pixels, $\Delta\lambda$ is the spectral 
dispersion in \AA/pixel, and EW represents the equivalent width of the line. The error in 
C$(H\beta)$ was computed by formally propagating the uncertainties of the line strengths, $I(\lambda)$

\begin{equation}
\sigma_{c(H \beta)}=\sigma_{I_{\lambda}/I_{H\beta}}
\left[\frac{1}{ln(10)[f(\lambda)-f(H\beta)](I_{\lambda}/I_{H\beta})}\right] .
\end{equation} 

Reddening-corrected emission-line intensities, normalized to H$\beta$ are presented in Table~\ref{table3} 
for the objects observed along with their extinction coefficient C(H$\beta$). The errors in the fluxes quoted 
in Table~\ref{table3} include the contribution of the uncertainty in C(H$\beta$) as well as of the uncertainty 
in the instrument response calibration; systematic errors were not included. 

\subsubsection{Nebular analysis}

The electron density, n$_{e}$, was calculated from the [SII]$\lambda\lambda$6717/6731 ratio using the 
prescriptions of the five-level atom FIVEL program available in the IRAF {\it nebular} package (Shaw \& Dufour \cite{sd94}). 
The electron density values range from $\le$ 100 cm$^{-3}$, obtained for nearly one half of the sample, to 
values as high as 417 cm$^{-3}$ for FCC~33. 

We could not detect the temperature-sensitive line [OIII]$\lambda$4363 in any of 
our spectra, thus were unable to obtain direct oxygen abundance measurements. Different 
empirical and theoretical metallicity calibrations have been calculated over the past two decades as an 
alternative to estimating oxygen abundances without any direct electron temperature measurements 
(e.g. Pagel et al. \cite{pag79}; Alloin et al. \cite{a79}; Edmunds \& Pagel \cite{ep84}; McGaugh \cite{mg91}; 
Vilchez \& Esteban \cite{ve96}; Pettini \& Pagel \cite{pp04}; Pilyugin et al. \cite{p06}; Pilyugin, Vilchez
\& Thuan \cite{pvt10}). These calibrations are referred to as strong-line methods since they are based on 
strong emission-line ratios (e.g. [OII]$\lambda$3727; [OIII]$\lambda\lambda$4959, 5007; [NII]$\lambda$6584; 
[SII]$\lambda\lambda$6717, 6731; [SIII]$\lambda\lambda$9069, 9532). 

For the determination of abundances in this paper, we used three methods: (i) the calibration of R23 
[R23 $\equiv$ ([OII]$\lambda$3727 + [OIII]$\lambda\lambda$4959,5007)/H$\beta $], via the theoretical models 
of McGaugh \cite{mg91} (hereafter M91) and analytical expressions in Kobulnicky et al. (\cite{k99}); 
(ii) the empirical calibration of the parameter N2 (N2 $\equiv$ [NII]$\lambda$6584/H$\alpha$) proposed 
by Pettini \& Pagel (\cite{pp04}; hereafter PP04); and (iii) the empirical calibration of the strong 
emission lines of oxygen, nitrogen, and sulfur (Pilyugin, Vilchez \& Thuan \cite{pvt10}; hereafter PVT10). 

The relation between 12 + log(O/H) and log(R23) is double valued, showing an ambiguity 
between the high and low abundance branches and a turnover region centered around 
12 + log(O/H) $\sim 8.2$ (e.g. M91; Miller \& Hodge \cite{mh96}; Olofsson \cite{o97}). Efforts have 
been made to refine the calibration of the R23 parameter. One of the most popular methods is the calibration 
by M91, which is based on photoionization models. As commonly accepted, we assumed that all the galaxies with 
log([NII]/[OII]) $< -1.0$ correspond to the lower oxygen abundance branch of this calibration, and conversely, 
the high abundance prediction of R23 correspond to objects showing log([NII]/[OII]) $\geq -1.0$ (e.g., 
McGaugh 1994; van Zee et al. 1998b; Contini et al. 2002). The typical uncertainties quoted for R23 empirical 
abundances are $\sim 0.10 - 0.20$ dex. In the turnover region, corresponding to log(R23) $\ge$ 0.9, the abundance 
predictions become more uncertain. Thus, in this work, we derived a conservative oxygen abundance of 
12 + log(O/H) = 8.20 using the $R23$ method for those objects in the sample presenting log (R23) $\ge$ 0.9. 

The N2 parameter (N2 $\equiv$ [NII]$\lambda$6584/H$\alpha$) offers several advantages, because 
it involves easily measurable lines that are available for a wide redshift range (up to z $\sim$ 2.5). 
The N2 versus O/H relation seems monotonic and the [NII]/H$\alpha $ ratio does not depend on reddening
correction or flux calibration. The drawbacks are that the [NII] lines can be affected by other 
excitation sources (see Van Zee et al. \cite{vz98}). In addition, N2 is sensitive to ionization conditions 
and relative N/O abundance variations. PP04 revised the N2 index, using 137 extragalactic HII regions. 
The uncertainty in the metallicity determination based on this calibration amounts to $\sim$ 0.3 dex.

PVT10 proposed some improved empirical calibrations for the determination of metallicity 
in HII regions. The authors refer to these calibrations as ONS calibrations since they are based 
on the strong emission lines of oxygen, nitrogen, and sulfur. 
They use both N2/R2 (=[NII]$\lambda\lambda$6548,6584/[OII]$\lambda$3727) and S2/R2 
(=[SII]$\lambda\lambda$6717, 6731/[OII]$\lambda$3727) ratios as temperature and metallicity indices. 
PVT10 found that the rms difference between oxygen abundance derived from the ONS calibration and an 
electron temperature based oxygen abundance is within 0.075 dex. 

For each galaxy studied in our paper, we adopted an average oxygen abundance obtained from the average of
the three (O/H) values obtained from the three calibrators R23, N2, and ONS. However, in the computation of 
the R23 abundance we follow the [NII]/[OII] criterion, as indicated above, to help us select between the 
upper or lower branch values. In Table~\ref{table3}, we include in parenthesis the rejected R23 values. 

Table~\ref{table3} includes the physical conditions, oxygen abundances, and their corresponding errors 
for our sample galaxies in Fornax and Hydra clusters.

%______________________________________________________________

\section{Discussion}
\label{discussion}

We now discuss the results obtained for individual objects based on our Gemini and VLT observations. 
We compared and complemented our findings with existing data from the literature. We consider the morphological 
and chemical analysis of our galaxies as well as the implications of our results within the framework of 
the fundamental relations governing galaxy evolution. 
 
\subsection{Morphological analysis}

\subsubsection{Fornax cluster}

FCC~32 is our second brightest Fornax object with a dense nucleus in all $K^\prime$, $R$, and net 
$H\alpha$ images. The poor conditions of the Gemini observations prevented us from resolving any details in 
the $R$ and $H\alpha$ images, but the VLT $K^\prime$ image observed in good seeing ($0.5\arcsec$) conditions 
unveiled about 6 off-center regions (see the middle image in Figure~\ref{fig2} in which about five regions 
appear as extended knots and one has a stellar appearance (the western-most one at $\alpha=$ 03:24:52.52, 
$\delta=$ -35:26:07.4 at J2000). Both $K^\prime$ and $R$ profiles are convex in the outer part and do not resemble 
a sech profile, while the colour is reddish $R-K^\prime\sim3$ mag. All these characteristics suggest a non-dwarf like structure 
for the host of FCC~32. Michielsen et al. (\cite{mic04}) observed this galaxy in $H\alpha$ and $R$ with the VLT in very 
good seeing, unveiling a bright core with a few bright knots associated with emission clouds similar to our findings, 
and classified this object as a non-nucleated dwarf elliptical (dE2). The authors found a total $R$ 
apparent magnitude of 14.52 mag, which matches very closely our Gemini findings $m_{TR}=14.50$ mag. These 
results do not support the findings of Drinkwater et al. (\cite{dri01a}), who classified this object as a BCD 
based on the detected $H\alpha$ emission observed with the UK Schmidt 1.2m telescope. Our net $H\alpha$ image 
also shows an important emission source in this galaxy. 

FCC~33 is our brightest Fornax object and shows a very elliptical-like morphology. Our Gemini $R$ and $H\alpha$ 
images show a bright nucleus with two extensions to the east and west. The $0.5\arcsec$ FWHM VLT $K^\prime$ 
image shows a very bright compact central nucleus at $\alpha=$ 03:24:58.51, $\delta=$ -37:00:35.4 (J2000) with many 
patchy regions distributed mostly along the east-west direction. Two arm structures are clearly apparent 
towards the east (winding north) and west (appearing to wind south), as can be seen in the right plot in 
Figure~\ref{fig2}. Both $K^\prime$ and $R$ radial profiles do not appear to conform to a sech profile fit, 
while the colour is reddish ($R-K^\prime\sim3$ mag decreasing slightly from centre to outer regions). All these findings  
suggest that FCC~33 has a spiral structure and is not a canonical dwarf galaxy. These results are not in the line  
with the findings of Drinkwater et al. (\cite{dri01a}), who classified this object as a BCD based on the spectroscopy 
taken with the UK Schmidt telescope. The total absolute magnitude inferred from our photometry, for the adopted 
distance modulus of Fornax, seems consistent with this suggestion.  

FCC~35 is our third brightest Fornax object showing some patchy star-like regions over its surface in 
both $R$ and $K^\prime$. The net $H\alpha$ image shows a very bright core surrounded by two neighboring patches 
to the north and other faint two to the south-east and south-west. The very compact relatively central 
nucleus (located slightly to the North of our centre fit along the principal axis) could be detected in 
both $R$ and net $H\alpha$ images at $\alpha=$ 03:25:04.21, $\delta=$ -36:55:35.6 (J2000), this structure being 
much fainter in $K^\prime$. The $K^\prime$ profile could be fitted with a sech plus a Gaussian. The $R-K^\prime$ colour 
varies between 1.5 and 3 mag, being blue at centre and redder at the outside. All these properties suggest that FCC~35
could be a BCD. These results agree with the findings of Drinkwater et al. (\cite{dri01a}), who classified 
this object as a BCD based on the $H\alpha$ emission found in the UK Schmidt spectra, and also
with the findings of Schroder, Drinkwater \& Richter (\cite{sch01}) who suggested a BCD classification. 

FCC~98 was not detected by the VLT in our $K^\prime$ band, being barely detected by Gemini in $R$ and $H\alpha$, 
but cannot be seen in the net $H\alpha$ image. We could not find any other observations of this galaxy in the 
literature.  

FCC~102 is our faintest Fornax galaxy and is relatively small, its $K^\prime$ and $R$ profiles being closely 
fitted by a sech model, this object having a blueish $R-K^\prime\sim2$ mag color. The pure $H\alpha$ image 
shows two compact features with a brighter elongated knot towards the south-east at $\alpha=$ 03:32:10.89, 
$\delta=$ -36:13:04.8 (J2000) and a fainter and more compact one towards the north-west at $\alpha=$ 03:32:10.34, 
$\delta=$ -36:12:59.8 (J2000). All these results suggests that FCC~102 could be a dI galaxy and agree with 
the findings of Drinkwater et al. (\cite{dri01a}), who detected $H\alpha$ emission, as well as those of Schroder, 
Drinkwater \& Richter (\cite{sch01}) who found HI emission in the Parkes radio data, both of these findings 
suggesting star formation activity. 

FCC~120 is a relatively faint object with a $K^\prime$ profile modeled well by a sech with a $R-K^\prime\sim2$ mag 
blueish colour indicative of a dI galaxy. The $K^\prime$ VLT $0.5\arcsec$ seeing allowed us to resolve a few regions 
(possibly bright star clusters), many patchy details being also visible in the $R$ image taken in poorer $4\times4$ 
binning at Gemini. These findings are similar to those obtained by Schroder, Drinkwater \& Richter (\cite{sch01}),
who detected HI emission consistent with star formation activity, and with the results of Seth et al. 
(\cite{set04}), who detected 12 star clusters in this galaxy using the HST, some visible as knots in our images. 

FCC~128 is faint and relatively small, having a $K^\prime$ profile closely modeled by a sech law and the $R-K^\prime\sim2$ mag 
blueish colour resembling that of a dI. The net $H\alpha$ image does not show any particular feature. The very
good VLT HAWK $0.4\arcsec$ seeing allowed us to resolve one starlike source and to identify an additional pair of patches 
close to the center of this galaxy. The results agree with those of Schroder, Drinkwater \& Richter (\cite{sch01}) who 
found HI emission. 

FCC~129 was not detected in $K^\prime$. The galaxy was barely detected in $R$, showing some fuzzy structure 
closed to its centre at $\alpha=$ 03:34:07.89, $\delta=$ -36:04:12.7 (J2000). Its net $H\alpha$ image does not 
show any emission. We could not find any other observations of this galaxy in the literature.  

FCC~130 was not detected in $K^\prime$, being also barely visible in $R$ and showing no net $H\alpha$ emission. 
Four structures close to the centre of the $R$ image are possibly associated with this galaxy, three of them being 
colinear along the east-west direction and another one visible to the south. At least two of these structures appear 
star-like ($FWHM=0.4-0.5\arcsec$). We could not find any previous observations of this galaxy in the literature.  

FCCB~905 was detected in both visible and NIR. Two very closed patches are visible in $R$ near the centre, 
and three patches are visible in the $H\alpha$ and net $H\alpha$ images at $\alpha=$ 03:33:57.24, $\delta=$ -34:36:42.3 
(J2000), the brightest being the most central, and the other two being fainter at $\alpha=$ 03:33:57.39, $\delta=$ -34:36:39.4 (J2000) 
and $\alpha=$ 03:33:57.15, $\delta=$ -34:36:46.5 (J2000). Interestingly, a fuzzy compact and very small elliptical patch 
could be detected in $K^\prime$ eastward and very close to the two north patches at $\alpha=$ 03:33:57.57, 
$\delta=$ -34:36:42.6 (J2000) but this object (possibly a background elliptical galaxy?) is undetected 
in our Gemini imaging in all bands. After removing this unknown object, the $K^\prime$ profile of FCCB~905 could be modeled 
with a sech plus Gaussian. The $R-K^\prime\sim2$ mag colour is blueish, thus FCCB~905 could be a BCD. These results agree with 
Drinkwater et al. (\cite{dri01a}) who found $H\alpha$ emission suggesting a BCD classification for this object, previously
identified (wrongly) with a background galaxy by Ferguson (\cite{fer89}). 

\subsubsection{Hydra cluster}

H1031-2818 represents by far our brightest Hydra object, showing a very bright core in $R$ and $H\alpha$ but 
presenting no net $H\alpha$ emission. Both $K^\prime$ and -to a lesser extent- $R$ images denote the appearance 
of some spiral structure (see the second zoomed $K^\prime$ image in Figure~\ref{fig2}). 
The surface profile in $K^\prime$ could be fitted by a sech model, but the one in $R$ seems more linear, while 
the colour is $R-K^\prime\sim2.5$ mag. Duc et al. (\cite{duc99}) detected this galaxy in HI. They also imaged it in $B$ 
and $I$ bands with the Danish 1.5m and in $K^\prime$ band with the ESO/MPG 2.2m telescope, finding exponential-like 
surface brightness profiles, similar to our findings. In the $K^\prime$ band, these authors found a total magnitude 
$m_K=12.69$ mag, which is close to -but fainter than- our VLT value $m_{TK}=12.27$ mag. 

H1031-2632 has a relatively faint nucleus in all bands, being surrounded by some fuzzy irregular envelope 
(see its $R$ and $H\alpha$ images in Figure~\ref{fig1}. 
Its $K^\prime$ profile could be well fitted with a sech plus a Gaussian, and the color $R-K^\prime\sim1.5-2$ mag seems 
blueish, so these suggest that this object is a BCD. Duc et al. (\cite{duc99}) imaged this object in $B$ and $I$ with 
the Danish telescope, finding surface brightness profiles almost exponential, similar to our Gemini results 
in $R$. These authors could not detect this object in $K^\prime$. 

H1032-2638 is an elliptical object, quite bright in $R$ but faint in $K^\prime$. Thanks to the good $0.9\arcsec$ 
Gemini seeing in $R$ band, this object clearly shows at its centre two bright stellar structures (visible 
also in $H\alpha$ and $K^\prime$) and two closer fainter unresolved features, reminiscent of a cross lensing effect. 
To fit the surface photometry, we removed the two brighter stellar-like structure via IMEDIT. Its surface 
brightness profile in $K^\prime$ could be fitted by a sech, while the color profile $R-K^\prime\sim1-2$ mag is very blue,
so that H1032-2638 could be a dI. Duc et al. (\cite{duc99}) derived $B$ and $I$ exponential profiles that flatten 
near the galaxy centre (i.e., similar to our sech law) but could not detect this object in $K^\prime$. 

H1034-2758 has a very compact central elliptical core that is clearly visible in $R$ and $H\alpha$ endowed in a 
fuzzy elliptic envelope oriented in the opposite direction of the core and showing some diffuse patches located mostly 
to the south and west. The core is also visible in $K^\prime$ but the envelope is less pronounced. Surface 
profiles in $K^\prime$ and $R$ are similar and could not be fitted with either a sech or Gaussian. The colour profile 
is blueish, $R-K^\prime\sim2$ mag. This object was also observed in $B$ and $I$ by Duc et al. (\cite{duc99}) who
showed light profiles similar to ours, and the authors also detected this objects in $K^\prime$ but their
reported magnitude $m_K=14.61$ mag is fainter than our VLT $m_{TK}=13.30$ mag. 

H1035-2756 presents a very strange shape for a galaxy: in both $R$ and $H\alpha$ frames it clearly shows a circular 
ring structure reminiscent of a planetary nebula, showing a light deficit clearly visible towards the centre and 
one bright non-stellar compact patch (as compared to field stars measured in poor $\sim2\arcsec$ seeing) 
located slightly to the east plus a few other fainter fuzzy patches appearing embedded in the ring envelope. 
The $K^\prime$ image does not show the envelope but only the brightest patch embedded in diffuse emission. 
Consequently, the $R$ light profile for this object winds twice, but the $K^\prime$ profile could be fit with 
a sech plus a Gaussian. The colour is blueish, $R-K^\prime\sim1-2$ mag, so this object could be a BCD candidate, 
although its ring in visible light remains to be explained. A round fuzzy patch could be observed in all bands 
located about 1 arcmin to the south-east of H1035-2756 ($\alpha=$ 10:37:414.41, $\delta=$ -28:12:56.6 at J2000) 
having an apparent diameter of about 10 arcsec. NED does not contain data for this object, so we can speculate 
that it is a new dwarf galaxy in Hydra, possibly a dI. Duc et al. (\cite{duc99}) also observed this peculiar object 
in the optical, and their profiles are very similar to our $R$ profile. 

H1035-2605 is our first brightest object in $R$ and the second one in $K^\prime$. It shows a very elliptical 
morphology, being a very long object with a very compact central core visible in all bands and some 
light deficit line centered along its north-eastern side resembling a dust lane. Its $K^\prime$ image shows 
a very extended and thin linear structure. Its light profile in $K^\prime$ being quite convex and unable to be modeled 
with any of our dwarf models, while its colour profile decreases from a central very red color $R-K^\prime\sim3.5$ mag 
to some blueish color in the outskirts $R-K^\prime\sim1.5$ mag. 
Given all these findings, H1035-2605 resembles a spiral 
viewed edge-on. The light profiles of Duc et al. (\cite{duc99}) in both visible and NIR are similar to
ours, and their reported colour is red, similar to our findings. 

H1035-2502 is a very faint and small object showing a slightly brighter prolonged core in $R$ 
and $H\alpha$ while its $K^\prime$ image is very diffuse and extremely faint. Its $K^\prime$ profile can accurately 
modeled with a sech and its color profile looks blueish, $R-K^\prime\sim1.5$ mag, so this object might be a dI. 
This agrees with the findings of Duc et al. (\cite{duc99}), whose visible profiles are sech-like and who managed 
to detect the object in HI. 

H1035-2740 is our faintest Hydra object in both $K^\prime$ and $R$, having an elliptical morphology 
with a bright star-like non-central core and some patchy structure visible mostly in $R$ towards 
the north-eastern side. Its $K^\prime$ profile could be modeled with a sech and its colour profile is
blueish, $R-K^\prime\sim1.5$ mag, thus this could be a dI. This result agrees with the finding of Duc et al. 
(\cite{duc99}) that the visible light profiles flatten towards the centre and that HI was observed throughout 
the galaxy, although these authors could not detect this object in $K^\prime$. 

H1038-2733 is our second faintest Hydra object, which is very small and diffuse in both visible and NIR. 
Its $R$ morphology looks strange: it has an extended core similar to a horizontal portion of a ring 
surrounded by two faint arm-like structures that make the overall structure of the galaxy appear 
elliptical. The $K^\prime$ surface is extremely faint and the profile could be fitted with a sech, while 
the color profile is blueish, $R-K^\prime\sim1.5$ mag. Given these findings, we regard this object as a dI candidate. 
This result agrees well with the findings of Duc et al. (\cite{duc99}), whose light profiles in the optical 
flatten towards the centre and who found that some HI emission was possible associated with this object. 

H1038-2730 has a central elliptical core in $R$, appearing clearly surounded by about three tight 
arms (see Figure~\ref{fig1}). Its $K^\prime$ image also contains a small elliptical core embedded in an elliptical 
fainter envelope, better visible to the south-east. Its light profiles look more linear than dwarf-like, 
so we regard this object as a background spiral. This agrees with Duc et al. (\cite{duc99}) whose light 
profiles in visible are almost exponential, similar to our findings in both visible and NIR. These 
authors classify this object as an edge-on system with isophote twisting without any evidence of 
current star-formation and some HI line width intriguingly broad. 

\subsection{Chemical abundance analysis}

We discuss in this section our derived abundances and compare them with existing data from the literature. 
We adopt the standard uncertainty of $\pm0.2$ dex for our derived abundances using the bright line methods 
discussed above. 

\subsubsection{Fornax cluster}

For FCC~32, we measured seven emission lines, including the bright
lines $[O~II]\lambda 3727$, $H\beta$, and $[O~III]\lambda 5007$ from which we 
calculated 12+log(O/H) = 8.68 and 8.15 using the upper and lower predictions of R23, 
respectively. For this galaxy, R23 upper was selected since log([NII]/[OII]) $\geq$ -1.0. 
Using the $N2$ indicator, we obtained 12+log(O/H) = 8.60 and from the ONS calibration we 
derived 12+log(O/H) = 8.52 dex. The adopted oxygen abundance, taken as the average among 
the three methods, is 12+log(O/H) = 8.60 dex. This high abundance implies that it is an 
old object, in accordance with our morphological findings, thus FCC~32 is definitely not 
a typical star-forming dwarf galaxy. 

The spectrum of FCC~33 does not exhibit all the emission lines required for the application 
of the R23 and ONS 
calibrations. For this galaxy, we only used the $N2$ indicator from which we derived 
12+log(O/H) = 8.55 dex. This relatively high O/H value suggests that it is a quite evolved 
object, in accordance with our morphological results, thus FCC~33 does not appear to be a 
star-forming dwarf galaxy. 
  
FCC~35 is the galaxy with the highest signal-to-noise ratio spectrum of our Fornax sample, for 
which we measured 13 emission lines.  Using the $N2$ indicator, we derived a metallicity 
12+log(O/H) = 7.87 dex; the ONS calibration gives us 12+log(O/H) = 7.81, while 
12+log(O/H) = 8.20 was assumed from the R23 method since log(R23)$>0.9$. The 
adopted abundance, 12+log(O/H) = 8.00 dex, suggests that it is a young object, consistent with 
our morphological findings. 

FCC~905 has nine measurable emission lines in its spectrum, based on which we derived 
12+log(O/H) = 8.19 and 8.20 using the N2 indicator and R23 method, respectively; 
from the ONS calibration we calculated 12+log(O/H) = 8.09. The average O/H gives us 
12+log(O/H) = 8.16 for this object. 

Two objects from Fornax did not show any emission lines, namely FCC~102 and FCC~120. 
We could not find any previous oxygen abundance determination for our Fornax sample 
to compare with our findings. 

\subsubsection{Hydra cluster}

The spectrum of H1031-2632 does not contain the emission lines needed to derive 
the oxygen abundances via the N2 and ONS calibrations. We adopt 12+log(O/H) = 8.20 dex, as 
obtained from the R23 method because log(R23)$>0.9$. Duc et al. (\cite{duc01}) acquired 
spectra for this object using the ESO 3.6m telescope but could not reliably detect any 
emission line. 

H1031-2818 shows seven emission lines from which we derived 12+log(O/H) = 8.65 
dex using the N2 method and 12+log(O/H) = 8.59 from the ONS calibration. The R23
upper and lower branches provide 12+log(O/H) = 8.98 and 7.53 dex, respectively. 
However, in the computation of the final O/H for this galaxy, we consider only the 
high metallicity prediction of R23 since log([NII]/[OII])$\geq$ -1.0.  Finally, we 
adopt 12+log(O/H) = 8.78 for this object. Duc et al. (\cite{duc01}) detected 10 emission 
lines for this object, deriving an abundance 12+log(O/H) = 8.81 using the R23 upper 
branch and 8.82 using the N2 method, both values close to our findings. 

From the spectrum of H1035-2502, we measured nine emission lines.  We derived 
12+log(O/H) = 8.47 and 8.28 from the high and low metallicity predictions of 
R23. Using the ONS calibration, we find 12+log(O/H) = 7.99, while the N2 method provides 
12+log(O/H) = 8.19. We adopt 12+log(O/H) = 8.17, selecting the R23 lower branch since 
log([NII]/[OII])$< -1.0$ for this galaxy. 

H1035-2605 has seven emission lines with measured fluxes. We derived 
12+log(O/H) = 8.56 using the N2 indicator, finding 8.78 for the R23 upper branch 
and 7.93 for the R23 lower branch. Because this galaxy shows log([NII]/[OII])$\geq -1.0$, 
we use the R23 upper branch in the computation of the final abundance value. From the ONS 
calibration, we find 12+log(O/H) = 8.50. Finally, we adopt an abundance of 12+log(O/H) = 8.63. 
Duc et al. (\cite{duc01}) measured nine lines for this object, deriving 12+log(O/H) = 8.65 
from the $N2$ method and 8.54 using the R23 upper branch, which are in agreement with 
our results, within the errors. 

Both H1035-2740 and H1035-2756 were observed only over the blue spectral range. We were unable 
to use the N2 and ONS calibrations owing to the lack of the nitrogen and sulfur emission 
lines. According to the R23 method, we assumed 12+log(O/H) = 8.20 for both galaxies, since 
they show log(R23) $\geq$ 0.90 placing the two galaxies in the turnover region of the R23 
calibration (see Section 3.2.2 for details). For H1035-2740, Duc et al. (\cite{duc01}) found 
12+log(O/H) = 7.84 based on a few indicators. Using the R23 lower branch, the same authors 
derived 12+log(O/H) = 7.95 for H1035-2756. 

\subsection{Fundamental relations}

\subsubsection{The fundamental plane}

Using a sample of dIs from the LV having known accurate distances, 
Vaduvescu \& McCall (\cite{vad08}) derived the so-called dwarf fundamental plane (FP) linking 
$K^\prime$ photometrical parameters (the central surface brightness and the ``sech'' 
magnitude) to the $HI$ velocity linewidths $W_{20}$ derived via radio data. In this paper
we use data for 34 objects published there. McCall et al. (\cite{mcc11}) augmented the 
FP dataset with an additional 16 objects, enlarging the FP sample 
to 50 objects from the LV, refining the FP to the relation 
\noindent
%\begin{multline}
\begin{equation}
\label{FP}
M_{SK} = (-3.90 \pm 0.34) \cdot \log(W_{20}) + (0.78 \pm 0.06) \cdot \mu_{0K} - (24.47 \pm 1.49)
\end{equation}
%\end{multline}
\noindent
In Figure~\ref{fig4}, we plot with the dashed line the dI FP defined based on this sample of dIs
(solid circles), taking into account data to be published soon (McCall et al. \cite{mcc11}). 

A useful method for examining the possible effects of the environment on the evolution of star-forming 
galaxies is to compare the properties of field objects with those of galaxies located in dense clusters 
(e.g. Vilchez \cite{vil95}; Vaduvescu and McCall \cite{vad08}; Ellison et al. \cite{elli09}). 
In this sense, we probed the FP using a sample of star-forming dwarf galaxies (mostly BCDs and a few dIs) 
located in the Virgo cluster (Vaduvescu, Richer \& McCall \cite{vad06}) by assuming a common distance to 
Virgo $DM=30.62$ (Freedman et al. \cite{fre01}). We found that most dwarfs in Virgo lie on the FP defined 
by the field dIs, thus BCDs and dIs appear to be similar structurally and dynamically. 
The scatter of Virgo dwarfs is larger, possibly because of cluster depth effects and/or the action 
of some environmental effects. 

The depth and substructure of a cluster should both be taken into account when studying 
distance-dependent parameters such as the absolute magnitude of a galaxy. 
Some previous studies of Virgo pointed out clear evidence to the three-dimensional substructure of the cluster 
(e.g., Gavazzi et al. \cite{gav99}), so that Gavazzi et al. (\cite{gav05}) published a revised 
list including 355 late-type members of Virgo located in seven clouds (A, E, S, N - at 17 Mpc, 
B - at 23 Mpc, and W and M - the most distant clouds, at 32 Mpc). On the basis of this list, we correct 
the distance modulus for two dwarfs in our Virgo sample, that is VCC~24 and VCC~144 (located 
in clouds M and W, respectively), $DM=32.53$, while for the remaining objects we retain our 
formerly adopted distance modulus $DM=30.62$ from Freedman et al. (\cite{fre01}). 
In Figure~\ref{fig4}, we overlay with solid triangles our updated sample of star-forming 
dwarfs in the Virgo cluster. 
As one can observe, most Virgo objects, including VCC~24 and VCC~144, are located now 
very close to or on the FP defined by the LV dwarfs, confirming that the cluster has 
no major enviromental influences on our selected Virgo sample of star-forming dwarfs. 

Only three dwarfs from our Fornax sample have published $W_{20}$ data (Schroeder et al. \cite{sch01}), 
while all objects in our Hydra sample have radio data (McMahon \cite{mcm93}; Duc et al. \cite{duc99}). 
We include their $W_{20}$ values in Table~\ref{table2} together with their reference source. Among these 
objects, only two galaxies in Fornax (FCC~35, FCC~120) and seven galaxies in Hydra have luminosity profiles 
consistent with dI or BCD profiles. We plot these objects on the dwarf FP in Figure~\ref{fig4} using open 
squares for Fornax objects and open starred symbols for Hydra objects, labeling only their last four 
digit names (for the sake of clarity). As one can observe, only two dwarfs from Fornax and Hydra 
(FCC~120 and H1032-2638) appear to be consistent (within the scatter) with the FP fit, while the majority 
of Hydra objects are located well above, towards the upper-left side of the plot. 

We assessed the effect of an uncertainty in the distance for Hydra, for which we adopted the single 
distance found in the literature, namely $d = 46\pm5$ Mpc (Jensen et al. 1999). Nevertheless, two main 
findings advocate that no such a distance effect could be responsible for such a large shift of most Hydra 
objects in the FP. The first finding is that most Hydra objects do fit the other three metallicity plots 
(presented next). The second finding is that to fit the plane, most Hydra points need a vertical 
shift (in absolute magnitude) of about 3 mag below, which means some 15 Mpc shift in distance, which 
is about three times larger than the uncertainty admitted by Jensen et al. Given this, we suggest that the 
environment has some some kind of influence for Hydra, e.g. causing the relatively low internal (gas) rotation. 
We closely inspected the published HI profiles of these Hydra galaxies, discarding 
any abnormal effects. On the other axis of the plot, absolute M$_{SK}$ magnitudes were derived 
from a sech fitting and appear to follow the model, as we describe in the next section. 

To check for possible cluster environmental effects, we calculated the cluster-centric 
distances for the objects in both samples with respect to the cluster centres reported by NED. 
For Fornax, the NED centre coincides with Chandra centre, while for Hydra this is only $2\arcmin$ 
away from the ROSAT field centre. 
Fornax galaxies reside between $0.9\deg$ and $3.1\deg$ from the cluster centre, from about 
$\sim$ 0.3 to 1.1 Mpc (projected at a 20 Mpc cluster distance) compared with $\sim$ $3.2\deg$ 
apparent radius of the cluster (Ferguson, \cite{fer89}), which is about 1 Mpc. 
Hydra objects are located between $0.5\deg$ and $1.2\deg$ away from the cluster centre, 
namely from about 0.4 to 1.0 Mpc (projected at a 46 Mpc cluster distance) compared with 
$\sim$$1.5\deg$ radius of the cluster (Fitchett and Merritt, \cite{fit88}), which is about 
1.3 Mpc. 
The available X-ray maps cover only $0.4\deg$ from the centre of Fornax and $0.3\deg$ for Hydra, 
being too small to include any object from our samples located in the outer regions of 
the two clusters. 

We analysed all the objects observed by Duc et al. (\cite{duc99}; \cite{duc01}), finding three more HI-rich 
dwarf galaxies with NIR, spectral, and radio data, in addition to another galaxy having only NIR and spectral data. 
We include these objects in Table~\ref{table4} and compare them with our samples. Using cross symbols, we 
overlay these objects in Figure~\ref{fig4}, adopting for them total magnitudes instead of sech magnitudes. 
Three galaxies are located close to the FP, and one object, H1038-2733, resides to the left. 

More observations are needed for a larger sample of star-forming dwarf galaxies across the entire 
extent of the Hydra and Fornax clusters to study in great depth the environmental hypothesis suggested 
by the above results. Since several of the currently invoked environmental effects are strongly 
related to the density and spatial distribution of the intergalactic medium (Boselli \& Gavazzi \cite{bos06}), 
a first test should be done using the X-ray distribution of these clusters. Unfortunately, we could not carry 
out this exercise based on the present small available X-ray coverage of the two clusters.

\subsubsection{Luminosity - metallicity relation}

The galaxy NIR emission is believed to be a more reliable tracer than visible light of the for the stellar 
mass of the old (more than 3 Gyr) populations in star-forming galaxies (Vaduvescu et al. \cite{vad05}), 
hence $K^\prime$ band can be successfully used to help us measure the stellar mass of a galaxy. Moreover,
NIR light and specifically that of the $K^\prime$ band is a more robust probe than any visible bands for
studying the chemical evolution 
of star-forming dwarf galaxies, particularly the luminosity - metallicity ($L-Z$) relation (Vaduvescu, 
McCall \& Richer \cite{vad07}; Saviane et al. \cite{sav08}; Guseva et al. \cite{gus09}). On the basis of 
a sample of 25 dIs and 14 BCDs from the LV and Virgo cluster, Vaduvescu, McCall \& Richer (\cite{vad07}) 
derived the $L_K-Z$ relations for dIs and BCDs (plotted with a dotted and a dashed line, respectively, 
in Figure~\ref{fig5}). These relations have a goodness of the fit of 0.10 and 0.11 dex, respectively, 
and the BCDs $L_K-Z$ relation is slightly steeper than the one for dIs. 

In the Figure~\ref{fig5}, we plot the data for four Fornax dwarfs and six Hydra objects from our sample using available data. 
Using cross symbols, we overlay the other four Hydra galaxies from Table~\ref{table4} (Duc et al. \cite{duc99}; 
Duc et al. \cite{duc01}), adopting for them total magnitudes instead of sech magnitudes. One can observe that most 
points from our data-set appear to be consistent with these relations (within the errors), with only 2-3 outliers, among 
which there is only one object from our samples, namely FCC~35. The average uncertainty in metallicities derived with 
bright line methods is $\sim$ 0.2 dex, so that each Fornax and Hydra data point on the diagram should be regarded 
with caution and as being subject to a possible vertical shift by this amount.

\subsubsection{Mass - metallicity relation}

To examine the relation between galaxy mass and metallicity, one should go beyond stellar mass 
to assess baryonic mass, which is defined to consist only of the mass contained in stars and gas. 
This quantity is significant 
especially in a star-forming galaxy where the gas mass can be a substantial part of the total mass budget. 
Radio observations of the $HI$ hydrogen species is known to provide such data based on the flux of the 
21cm observations, but this line is rarely detected for faint objects and requires large radio dishes. 
Only one Fornax object and five Hydra objects from our sample have the requested $HI$ data available, from 
which we calculated gas masses using the same relations as in Vaduvescu, McCall \& Richer (\cite{vad07}). 
We included in Table~\ref{table2} the logarithm of the stellar mass, gas mass, and baryonic mass, 
which are all given in Solar units. 
In Figure~\ref{fig6}, we plot our samples, together with the comparison samples for Virgo and LV. 
With cross symbols, we overlay three other objects with data from Duc et al. (\cite{duc99}; \cite{duc01}). 
There is a conspicuous outlier in our sample, namely the Hydra galaxy H1031-2818, which has been 
suggested to be a possible spiral galaxy candidate. 

Star-forming dwarfs (dIs and BCDs) located in isolation appear to follow a different mass-metallicity 
relation than similar galaxies located in the Virgo cluster (Vaduvescu, McCall \& Richer, \cite{vad07}), thus 
it is very important to study this relation further taking into account our LV, Virgo, and Hydra samples. 
On the basis of 21 star-forming dwarfs in the LV (17 dIs and 4 BCDs), we derived the following 
mass-metallicity fit using a classical fit that minimizes the standard error of a 
linear distribution (LINEST function in Linux/OpenOffice)
\begin{equation}
12+\log(\rm{O/H}) = (5.80 \pm 0.47) + (0.25 \pm 0.06) \log(M_{bary}) .
\end{equation}
\noindent
On the basis of 14 star-forming dwarfs in Virgo cluster (3 dIs and 11 BCDs), we derive the 
mass-metallicity relation 
\begin{equation}
12+\log(\rm{O/H}) = (2.79 \pm 0.83) + (0.64 \pm 0.10) \log(M_{bary}) .
\end{equation}
\noindent
Using 8 star-forming dwarf candidates with available data in Hydra cluster (our 5 galaxies 
plus 3 objects from Duc et al. \cite{duc01}), we derive the mass-metallicity relation 
\begin{equation}
12+\log(\rm{O/H}) = (1.09 \pm 1.89) + (0.78 \pm 0.21) \log(M_{bary}) .
\end{equation}
\noindent
Unfortunately we have only one Fornax object with available data, so we cannot check the Fornax 
mass-metallicity relation. In Figure~\ref{fig6}, we draw with a solid line the LV fit (dIs and BCDs), 
with a dotted line the Virgo fit, and with a dashed line the Hydra fit.  

One can observe that the slope of the fit is shallower than for the LV sample (0.25), and higher for the 
Virgo sample 
(0.64) and the Hydra sample (0.78). This could suggest an environmental trend for star-forming dwarfs in the 
mass-metallicity relation, with more metal-rich objects favouring regions of higher galaxy overdensity, namely 
Virgo and Hydra clusters. This result is in the line with Petropoulou et al. (\cite{pet11}), who demonstrated 
that some star-forming dwarfs located at higher densities in the Hercules cluster tend to be more metal rich 
than the lower density counterparts. 

Treated as separate entities, dIs and BCDs appear to follow different mass-metallicity relations, with BCDs 
following a steeper relation than dIs alone. This could be checked for both LV and Virgo samples, 
and the result is similar to the luminosity-metallicity relation addressed by Vaduvescu, McCall 
\& Richer (\cite{vad07}). 

\subsubsection{Metallicity - gas fraction relation}

Lee et al. (\cite{lee03}) studied the closed box model of the chemical evolution of star-forming dwarf galaxies,
which predicts a linear correlation between metallicity and the logarithm of the logarithm (''log-log``) of the 
inverse of the gas fraction. We studied this relation for our (small) Fornax and Hydra sample of galaxies for which 
these data are available. We include in Table~\ref{table2} the gas fraction for the galaxies of the two cluster samples, 
and in Figure~\ref{fig7} we plot with a dotted line the metallicity - gas fraction relation, derived in Lee et al. \cite{lee03}, 
as well as the points of our published LV and Virgo samples. With cross symbols, we add three other Hydra 
galaxies with data available from Duc et al. (\cite{duc99}; \cite{duc01}), which all match the relation quite well. 
Although our two Fornax and Hydra samples are not statistically significant, one can see that most objects appear to be 
described by this relation. This suggests that the chemical evolution of star-forming galaxies in Fornax and Hydra 
is consistent with the theoretical expectations for a closed box model, thus seems compatible with a (relatively) 
unperturbed evolution.

%______________________________________________________________

\section{Conclusions}
\label{conclusions}

Two samples of ten star-forming galaxy candidates in Fornax cluster and ten objects in Hydra cluster have 
been selected based on data available in the literature to study their physical and chemical properties, 
as part of a larger project aimed to begin the study of star-forming galaxies in nearby clusters using 8-10m 
class telescopes. For these two samples, we acquired $H\alpha$ and $R$ imaging using Gemini/GMOS and $K^\prime$ 
imaging using the VLT/HAWK-I. Based on the $H\alpha$ images, we selected two sub-samples of 6 objects in 
Fornax and 6 objects in Hydra to be observed spectroscopicaly with Gemini/GMOS. 

Taking into account the $R$, $H\alpha$, and $K^\prime$ imaging, we studied the morphology of the two samples, 
finding five objects in Fornax and six objects in Hydra that have a structure compatible with a star-forming 
dwarf (dI or BCD-like), while about four objects are probably spiral galaxies and another three objects are 
virtually undetected. 
Although we aimed to detect the faint [OIII]$\lambda$4363 line to derive accurate direct $T_e$ oxygen abundances, 
the poor weather conditions and the large distance of Hydra precluded us from detecting this line in the observed 
galaxies. Taking into account the emission lines measured in the Gemini/GMOS specra, 
we derived metallicities for ten star-forming galaxy candidates (four in Fornax and six in Hydra) 
corresponding to abundances in the range 8.00 $\le$ 12+log(O/H) $\le$ 8.78. 

Based on the derived physical and chemical data, we studied several fundamental relations for a small 
sample of star-forming galaxies with available data in Fornax and Hydra. 

The luminosity - metallicity and the metallicity - gas fraction relations found for star-forming dwarfs 
in the LV and Virgo appear to be followed by our Fornax and Hydra sample of galaxies having available radio 
data. These results suggest that the chemical evolution of these galaxies in both clusters appears consistent 
with the predictions of a closed box model. 

Star-forming dwarfs (dIs and BCDs) in different environments appear to follow different mass-metallicity 
relations, with more metal-rich objects tending to occupy regions of higher galaxy overdensity, namely the 
Virgo and Hydra clusters. 
Treated separatelly, dIs and BCDs also appear to define different mass-metallicity relations, with 
BCDs following a steeper relation than dIs fitted alone, a result similar to the luminosity-metallicity 
relation. 

Only two cluster objects appear confined to the dwarf FP, which does not seem to hold for most Hydra 
objects. This suggests that the cluster environment has some effect in Hydra, while the Fornax available 
sample was too small to probe the FP. 

Based on our observations and some scarce data available in the literature (especially in the radio), 
we have identified some research avenues to studying star-forming galaxies in clusters. To confirm our 
results, a concertated effort to acquire 21-cm observations should be conducted in conjunction 
with NIR imaging and spectroscopic research of nearby clusters, so that most physical and chemical 
data could be available for studying galaxy evolution in clusters. 

%______________________________________________________________

\begin{acknowledgements}
O.V. thanks to the Chilean TACs for the time allocation at Gemini South (programmes GS-2007B-Q-53 
and GS-2008A-Q-49) and ESO VLT (382.A-0409A). Acknowledgements are due to Prof. Marshall McCall who 
suggested some important insights in the cluster and sample selection and to Dr. Henry Lee for joining 
the Gemini observing proposals. 
C.K., as a Humboldt Fellow, acknowledges support from the Alexander von Humboldt Foundation, Germany. 
J.M.V. and C.K. acknowledge support by projects: AYA2007-67965-C03-02, AYA2010-21887-C04-01 and 
Consolider-Ingenio 2010 CSD2006-00070 ``First Science with GTC'', of the spanish MICINN. 
This research has made use of the NASA/IPAC Extragalactic Database (NED) which is operated by the Jet 
Propulsion Laboratory, California Institute of Technology, under contract with the National Aeronautics 
and Space Administration. We acknowledge the usage of the HyperLeda database (http://leda.univ-lyon1.fr - 
Paturel et al. 2003). Our work used IRAF, a software package distributed by the National Optical 
Astronomy Observatory, which is operated by the Association of Universities for Research in Astronomy 
(AURA) under cooperative agreement with the National Science Foundation. This research has made use of 
SAOImage DS9, developed by Smithsonian Astrophysical Observatory. 
This paper is based on observations obtained at the Gemini Observatory (Run IDs: GS-2007B-Q-53 and 
GS-2008A-Q-49), which is operated by the Association of Universities for Research in Astronomy, Inc., 
under a cooperative agreement with the NSF on behalf of the Gemini partnership: the National Science 
Foundation (United States), the Science and Technology Facilities Council (United Kingdom), the 
National Research Council (Canada), CONICYT (Chile), the Australian Research Council (Australia), 
Ministerio da Ciencia e Tecnologia (Brazil) and Ministerio de Ciencia, Tecnologia e Innovacion 
Productiva  (Argentina). Special thanks are due to the Gemini staff who acquired and took care of 
our data, specificaly Rodrigo Carrasco, Henry Lee, Jose Gallardo, Kathleen Labrie, Claudia Winge and 
Pablo Candia. This paper is also based on observations made with ESO Telescopes at the La Silla or 
Paranal Observatories under programme ID 382.A-0409. 
Special thanks are due to the refferee whose comments helped us to improve the paper.

\end{acknowledgements}

%______________________________________________________________

\clearpage

\begin{tiny}
\begin{table}[p]
\begin{center}
\caption{The log of observations acquired at Gemini South and VLT of the star-forming dwarf galaxies 
in Fornax and Hydra clusters. $H\alpha$ (Ha\_G0336) and continuum (r\_G0326) filters refer to pre-imaging 
and B600/480 and R400/780 refers to blue and red spectroscopy (Gemini). 
K filter refers to NIR imaging with the VLT. } 
\label{obs_log}
\begin{scriptsize}
\begin{tabular}{lrrrrr}
\hline
\hline
\noalign{\smallskip}
Galaxy & $\alpha$ (J2000) & $\delta$ (J2000) & Date (UT) &  Filter  & Exp (s) \\
\hline
\noalign{\smallskip}

 FCC 32     & 03:24:52.4 & -35:26:08 & Sep 13, 2007 & Ha\_G0336 &  540 \\ 
 ...        &     ...    &     ...   & Sep 13, 2007 & r\_G0326  &  180 \\ 
 ...        &     ...    &     ...   & Jan 12, 2008 & B600/480  & 1800 \\ 
 ...        &     ...    &     ...   & Jan 10, 2008 & R400/780  &  600 \\ 
 ...        &     ...    &     ...   & Sep 17, 2008 & K         &  720 \\ 
 FCC 33     & 03:24:58.4 & -37:00:34 & Sep 13, 2007 & Ha\_G0336 &  540 \\ 
 ...        &     ...    &     ...   & Sep 13, 2007 & r\_G0326  &  180 \\ 
 ...        &     ...    &     ...   & Jan 06, 2008 & B600/480  & 1800 \\ 
 ...        &     ...    &     ...   & Jan 06, 2008 & R400/780  &  600 \\ 
 ...        &     ...    &     ...   & Sep 17, 2008 & K         &  720 \\ 
 FCC 35     & 03:25:04.2 & -36:55:39 & Sep 13, 2007 & Ha\_G0336 &  540 \\ 
 ...        &     ...    &     ...   & Sep 13, 2007 & r\_G0326  &  180 \\ 
 ...        &     ...    &     ...   & Dec 28, 2007 & B600/480  &  600 \\ 
 ...        &     ...    &     ...   & Dec 28, 2007 & R400/780  & 1800 \\ 
 ...        &     ...    &     ...   & Sep 17, 2008 & K         &  720 \\ 
 FCC 98     & 03:31:39.2 & -36:16:35 & Sep 13, 2007 & Ha\_G0336 &  540 \\ 
 ...        &     ...    &     ...   & Sep 13, 2007 & r\_G0326  &  180 \\ 
 ...        &     ...    &     ...   & Sep 24, 2008 & K         &  720 \\ 
 FCC 102    & 03:32:10.7 & -36:13:15 & Sep 11, 2007 & Ha\_G0336 &  540 \\ 
 ...        &     ...    &     ...   & Sep 11, 2007 & r\_G0326  &  180 \\ 
 ...        &     ...    &     ...   & Jan 12, 2008 & B600/480  & 1800 \\ 
 ...        &     ...    &     ...   & Jan 16, 2008 & R400/780  &  600 \\ 
 ...        &     ...    &     ...   & Sep 17, 2008 & K         &  720 \\ 
 FCC 120    & 03:33:34.2 & -36:36:21 & Sep 11, 2007 & Ha\_G0336 &  540 \\ 
 ...        &     ...    &     ...   & Sep 11, 2007 & r\_G0326  &  180 \\ 
 ...        &     ...    &     ...   & Jan 06, 2008 & B600/480  & 1800 \\ 
 ...        &     ...    &     ...   & Jan 06, 2008 & R400/780  &  600 \\ 
 ...        &     ...    &     ...   & Sep 17, 2008 & K         &  720 \\ 
 FCC 128    & 03:34:07.1 & -36:27:57 & Sep 12, 2007 & Ha\_G0336 &  540 \\ 
 ...        &     ...    &     ...   & Sep 12, 2007 & r\_G0326  &  180 \\ 
 ...        &     ...    &     ...   & Sep 18, 2008 & K         &  720 \\ 
 FCC 129    & 03:34:07.7 & -36:04:11 & Sep 13, 2007 & Ha\_G0336 &  540 \\ 
 ...        &     ...    &     ...   & Sep 13, 2007 & r\_G0326  &  180 \\ 
 ...        &     ...    &     ...   & Sep 24, 2008 & K         &  720 \\ 
 FCC 130    & 03:34:09.2 & -35:31:00 & Sep 13, 2007 & Ha\_G0336 &  540 \\ 
 ...        &     ...    &     ...   & Sep 13, 2007 & r\_G0326  &  180 \\ 
 ...        &     ...    &     ...   & Sep 24, 2008 & K         &  720 \\ 
 FCCB 905   & 03:33:57.2 & -34:36:43 & Sep 12, 2007 & Ha\_G0336 &  540 \\ 
 ...        &     ...    &     ...   & Sep 12, 2007 & r\_G0326  &  180 \\ 
 ...        &     ...    &     ...   & Dec 29, 2007 & B600/480  & 1800 \\ 
 ...        &     ...    &     ...   & Dec 29, 2007 & R400/780  &  600 \\ 
 ...        &     ...    &     ...   & Sep 17, 2008 & K         &  720 \\ 
\\
\hline
\\
 H1031-2818 & 10:34:16.6 & -28:34:05 & Feb 03, 2008 & Ha\_G0336 & 1080 \\ 
 ...        &     ...    &     ...   & Feb 03, 2008 & r\_G0326  &  360 \\ 
 ...        &     ...    &     ...   & May 08, 2008 & B600/480  & 5400 \\ 
 ...        &     ...    &     ...   & Apr 26, 2008 & R400/780  & 1800 \\ 
 ...        &     ...    &     ...   & Nov 29, 2008 & K         &  720 \\ 
 H1031-2632 & 10:34:20.6 & -26:47:31 & Feb 04, 2008 & Ha\_G0336 & 1080 \\  
 ...        &     ...    &     ...   & Feb 04, 2008 & r\_G0326  &  360 \\ 
 ...        &     ...    &     ...   & Apr 13, 2008 & B600/480  & 5400 \\ 
 ...        &     ...    &     ...   & May 12, 2008 & R400/780  & 1800 \\ 
 ...        &     ...    &     ...   & Dec 1,  2008 & K         &  720 \\ 
 H1032-2638 & 10:34:40.5 & -26:54:33 & Feb 03, 2008 & Ha\_G0336 & 1080 \\  
 ...        &     ...    &     ...   & Feb 11, 2008 & r\_G0326  &  360 \\ 
 ...        &     ...    &     ...   & Dec 4,  2008 & K         &  720 \\ 
 H1034-2758 & 10:37:19.9 & -28:14:20 & Feb 08, 2008 & Ha\_G0336 & 1080 \\  
 ...        &     ...    &     ...   & Feb 08, 2008 & r\_G0326  &  360 \\ 
 ...        &     ...    &     ...   & Dec 4,  2008 & K         &  720 \\ 
 H1035-2756 & 10:37:38.6 & -28:12:25 & Feb 03, 2008 & Ha\_G0336 & 1080 \\  
 ...        &     ...    &     ...   & Feb 03, 2008 & r\_G0326  &  360 \\ 
 ...        &     ...    &     ...   & May 01, 2008 & B600/480  & 5400 \\ 
 ...        &     ...    &     ...   & Dec 4,  2008 & K         &  720 \\ 
 H1035-2605 & 10:37:41.0 & -26:20:55 & Feb 03, 2008 & Ha\_G0336 & 1080 \\  
 ...        &     ...    &     ...   & Feb 03, 2008 & r\_G0326  &  360 \\ 
 ...        &     ...    &     ...   & Apr 14, 2008 & B600/480  & 5400 \\ 
 ...        &     ...    &     ...   & Apr 29, 2008 & R400/780  & 1800 \\ 
 ...        &     ...    &     ...   & Nov 29, 2008 & K         &  720 \\ 
 H1035-2502 & 10:37:51.3 & -25:18:07 & Feb 03, 2008 & Ha\_G0336 & 1080 \\  
 ...        &     ...    &     ...   & Feb 03, 2008 & r\_G0326  &  360 \\ 
 ...        &     ...    &     ...   & May 12, 2008 & B600/480  & 5400 \\ 
 ...        &     ...    &     ...   & Apr 27, 2008 & R400/780  & 1800 \\ 
 ...        &     ...    &     ...   & Dec 4,  2008 & K         &  720 \\ 
 H1035-2740 & 10:38:11.9 & -27:56:14 & Feb 03, 2008 & Ha\_G0336 & 1080 \\  
 ...        &     ...    &     ...   & Feb 03, 2008 & r\_G0326  &  360 \\ 
 ...        &     ...    &     ...   & Apr 26, 2008 & B600/480  & 5400 \\ 
 ...        &     ...    &     ...   & Dec 8,  2008 & K         &  720 \\ 
 H1038-2733 & 10:40:26.6 & -27:48:52 & Feb 03, 2008 & Ha\_G0336 & 1080 \\  
 ...        &     ...    &     ...   & Feb 03, 2008 & r\_G0326  &  360 \\ 
 ...        &     ...    &     ...   & Dec 3,  2008 & K         &  720 \\ 
 H1038-2730 & 10:40:59.4 & -27:45:42 & Feb 04, 2008 & Ha\_G0336 & 1080 \\  
 ...        &     ...    &     ...   & Feb 04, 2008 & r\_G0326  &  360 \\ 
 ...        &     ...    &     ...   & Dec 8,  2008 & K         &  720 \\ 

\noalign{\smallskip}
\hline
\hline
\end{tabular}
\end{scriptsize}
\label{table1}
\end{center}
\end{table}
\end{tiny}

%______________________________________________________________

\clearpage

\begin{table}[!t]
\begin{center}
\caption{Physical parameters of the galaxies observed in Fornax and Hydra: (1) galaxy name; (2) ellipticity; (3) position angle (degrees); 
(4) total apparent magnitude in $K^\prime$; (5) sech magnitude in $K^\prime$; (6) central surface brightness in $K^\prime$ (mag/arcsec$^2$); 
(7) sech $K^\prime$ scale radius (arcsec); (8) radius (arcsec) at $K^\prime=22$ mag/arcsec$^2$; (9) total apparent magnitude in $R$, $m_{TR}$; 
(10) absolute sech $K^\prime$ magnitude (in paranthesis total $K^\prime$ magnitude); (11) hydrogen velocity linewidth at $20\%$ height, 
$W_{20}$ (km/s); (12) reference for $W_{20}$ [1: Schroeder et al. 2001 (Parkes); 2: Duc et al. 1999 (Nancay); 3: McMahon 1993 (VLA)]; 
(13) logarithm of galaxy stellar mass, log M$_*$ (M$_\odot$); (14) logarithm of galaxy gas mass, log M$_G$  (M$_\odot$); (15) logarithm 
of baryonic mass, log M$_B$ (M$_\odot$); (16) gas fraction $\mu$. 
} 
\begin{tabular}{lrrrrrrrrrrrrrrr}
\hline
\hline
\noalign{\smallskip}
$Galaxy$ & $e$ & $PA$ & $m_{TK}$ & $m_{SK}$ & $\mu_{OK}$ & $r_{OK}$ & $r_{22K}$ & $m_{TR}$ & $M_{SK}$ & $W_{20}$ & Ref & log M$_*$ & log M$_G$ & log M$_{bary}$ & $\mu$ \\
\hline
\noalign{\smallskip}

 FCC 32     &  0.3  &  -25  &  11.92  &   ---   &   ---   &   ---   &   ---   &  14.50  & (-19.59) &  --- & --- & (9.06) &   --- &   --- &   --- \\
 FCC 33     &  0.6  &  +80  &  10.60  &   ---   &   ---   &   ---   &   ---   &  13.56  & (-21.91) &   63 & (1) & (9.99) &  8.46 &  9.56 &  0.08 \\
 FCC 35     &  0.6  &  +15  &  13.11  &  12.41  &  19.51  &   9.3   &  27.77  &  14.97  &  -19.10  &   56 & (1) &  8.86  &   --- &   --- &   --- \\ 
 FCC 102    &  0.3  &  -45  &  14.37  &  14.22  &  20.61  &   6.7   &  13.10  &  15.73  &  -17.29  &  --- & --- &  8.14  &   --- &   --- &   --- \\ 
 FCC 120    &  0.5  &  +35  &  14.00  &  13.93  &  20.66  &   9.3   &  17.70  &  15.74  &  -17.58  &  141 & (1) &  8.26  &   --- &   --- &   --- \\ 
 FCC 128    &  0.2  &  +30  &  14.27  &  14.16  &  20.09  &   5.1   &  12.39  &  16.06  &  -17.35  &  --- & --- &  8.16  &   --- &   --- &   --- \\ 
 FCCB 905   &  0.3  &  +30  &  14.34  &  14.46  &  20.36  &   5.3   &  11.69  &  15.73  &  -17.05  &  --- & --- &  8.04  &   --- &   --- &   --- \\ 

\noalign{\smallskip}
\hline
\noalign{\smallskip}

 H1031-2818 &  0.3  &  +70  &  12.27  &  12.29  &  17.00  &   3.1   &  16.35  &  14.76  &  -21.02  &  102 & (2) &  9.63  &  8.51 &  9.66 &  0.07 \\
 H1031-2632 &  0.2  &  +20  &  14.40  &  14.40  &  20.37  &   5.2   &  11.26  &  15.79  &  -18.91  &   81 & (2) &  8.79  &  8.40 &  8.94 &  0.29 \\
 H1032-2638 &  0.5  &  +75  &  15.64  &  15.62  &  20.23  &   3.5   &   8.08  &  16.52  &  -17.69  &  128 & (3) &  8.30  &   --- &   --- &   --- \\
 H1034-2758 &  0.3  &  -50  &  13.30  &   ---   &   ---   &   ---   &   ---   &  15.00  & (-20.01) &  136 & (2) & (9.23) &  8.81 &  9.46 &  0.23 \\
 H1035-2756 &  0.2  &  -30  &  15.10  &  15.11  &  21.49  &   6.2   &   6.53  &  16.42  &  -18.20  &   94 & (2) &  8.50  &  8.95 &  9.08 &  0.74 \\
 H1035-2605 &  0.8  &  +25  &  12.88  &   ---   &   ---   &   ---   &   ---   &  15.29  & (-20.43) &   84 & (3) & (9.40) &  9.32 &  9.67 &  0.45 \\
 H1035-2502 &  0.1  &  +90  &  15.47  &  15.33  &  20.64  &   3.6   &   6.92  &  16.72  &  -17.98  &   88 & (2) &  8.42  &  8.53 &  8.78 &  0.57 \\
 H1035-2740 &  0.7  &  +35  &  15.77  &  15.53  &  20.28  &   4.8   &  10.87  &  16.85  &  -17.78  &   84 & (3) &  8.34  &   --- &   --- &   --- \\
 H1038-2733 &  0.5  &  +90  &  15.76  &  15.73  &  20.64  &   4.0   &   7.70  &  16.56  &  -17.58  &   67 & (2) &  8.26  &  8.85 &  8.95 &  0.80 \\
 H1038-2730 &  0.6  &  -55  &  14.59  &   ---   &   ---   &   ---   &   ---   &  16.10  & (-18.72) &  173 & (2) & (8.71) &  8.94 &  9.14 &  0.62 \\

\noalign{\smallskip}
\hline
\hline
\end{tabular}
\label{table2}
\end{center}
\end{table}

%______________________________________________________________

\clearpage

%\onecolumn
\begin{sidewaystable}
%\begin{tiny}
%\begin{table}
\caption{Reddening-corrected line fluxes relative to F(\mbox{H$\beta$})=1 for the sample of Fornax and Hydra galaxies}
\label{table3}
\begin{center}
\begin{tabular}{lcccccccccccc}
\hline\hline
\\
Wavelength      &  FCC 32           &  FCC 33          &  FCC 35            &  FCCB 905        &  H1031-2632      &  H1031-2818      &  H1035-2502      &  H1035-2605      &  H1035-2740      & H1035-2756       \\ 
\\
\hline
\\
3727 [O~II]     &  3.91 $\pm$ 0.26  &  1.26 $\pm$ 0.03 & 2.52 $\pm$ 0.13    &  5.73 $\pm$ 0.92 &  5.44 $\pm$ 1.18 &  1.51 $\pm$ 0.04 &  5.08 $\pm$ 1.61 &  3.00 $\pm$ 0.45 &  4.08 $\pm$ 0.66 &  3.61 $\pm$ 0.64 \\
3868 [Ne~III]   &   ---             &   ---            & 0.75 $\pm$ 0.07    &  ---             &  ---             &  ---             &   ---            &   ---            &   ---            &   ---            \\
3968H7+[Ne~III] &   ---             &   ---            & 0.31 $\pm$ 0.02    &  ---             &  ---             &  ---             &   ---            &   ---            &   ---            &   ---            \\
4100 H$\delta$  &   ---             &   ---            & 0.30 $\pm$ 0.02    &  ---             &  ---             &  ---             &   ---            &   ---            &   ---            &   ---            \\
4340 H$\gamma$  &   ---             &   ---            & 0.47 $\pm$ 0.01    &  0.48 $\pm$ 0.04 &  ---             &  ---             &  0.49 $\pm$ 0.10 &   ---            &  0.49 $\pm$ 0.05 &  0.47 $\pm$ 0.05 \\
4860 H$\beta$   &  1.00 $\pm$ 0.02  &  1.00 $\pm$ 0.01 & 1.00 $\pm$ 0.01    &  1.00 $\pm$ 0.03 &  1.00 $\pm$ 0.05 &  1.00 $\pm$ 0.01 &  1.00 $\pm$ 0.04 &  1.00 $\pm$ 0.04 &  1.00 $\pm$ 0.03 &  1.00 $\pm$ 0.07 \\  
4959 [O~III]    &   ---             &   ---            & 1.88 $\pm$ 0.02    &  0.66 $\pm$ 0.02 &  1.03 $\pm$ 0.10 &  ---             &  0.48 $\pm$ 0.03 &  ---             &  1.15 $\pm$ 0.04 & 1.05 $\pm$ 0.03   \\
5007 [O~III]    &  0.32 $\pm$ 0.02  &   ---            & 5.57 $\pm$ 0.05    &  1.96 $\pm$ 0.06 &  2.75 $\pm$ 0.21 &  0.24 $\pm$ 0.01 &  1.31 $\pm$ 0.07 &  0.46 $\pm$ 0.04 &  3.27 $\pm$ 0.10 &  3.20 $\pm$ 0.08 \\
6563 H$\alpha$  &  2.87 $\pm$ 0.10  &  2.89 $\pm$ 0.05 & 2.86 $\pm$ 0.10    &  2.86 $\pm$ 0.38 &  2.93 $\pm$ 0.27 &  2.87 $\pm$ 0.04 &  2.86 $\pm$ 0.70 &  2.95 $\pm$ 0.14 &  ---             &   ---            \\  
6584 [N~II]     &  0.85 $\pm$ 0.04  &  0.75 $\pm$ 0.01 & 0.036 $\pm$ 0.002  &  0.17 $\pm$ 0.03 &  ---             &  0.94 $\pm$ 0.01 &  0.18 $\pm$ 0.05 &  0.76 $\pm$ 0.04 &  ---             &   ---            \\ 
6678 He~I       &      ---          &   ---            & 0.027 $\pm$ 0.001  &  ---             &  ---             &  ---             &   ---            &  ---             &  ---             &   ---            \\ 
6717 [S~II]     &  0.82 $\pm$ 0.03  &  0.39 $\pm$ 0.01 & 0.104 $\pm$ 0.004  &  0.39 $\pm$ 0.06 &  0.77 $\pm$ 0.09 &  0.68 $\pm$ 0.01 &  0.44 $\pm$ 0.13 &  0.76 $\pm$ 0.04 &  ---             &   ---            \\ 
6731 [S~II]     &  0.63  $\pm$ 0.03 &  0.36 $\pm$ 0.01 & 0.072 $\pm$ 0.003  &  0.25 $\pm$ 0.04 &  0.62 $\pm$ 0.07 &  0.44 $\pm$ 0.01 &  0.29 $\pm$ 0.09 &  0.53 $\pm$ 0.03 &  ---             &   ---            \\ 
\\
\hline
\\
$C_{H\beta}$    
   &  0.01 $\pm$ 0.04  &  0.18 $\pm$ 0.02 &  0.08 $\pm$ 0.05   &  0.51 $\pm$ 0.19 &  0.74 $\pm$ 0.08 &  0.11 $\pm$ 0.01 &  0.91 $\pm$ 0.41 &  0.78 $\pm$ 0.05 &  0.68 $\pm$ 0.20 &  0.18 $\pm$ 0.21 \\
$n_{e} (cm^{-3})$  
  &  116              &  417             &  $<$ 100           & $<$ 100          &  198             &  $<$ 100         &  $<$ 100         &  $<$ 100         &  ---             &  ---             \\
12+log(O/H)$_{N2}$\footnote{O/H derived from Petini \& Pagel \cite{pp04} - N2 method}
  &  8.60             &  8.55            &  7.87              &  8.19            &  ---             &  8.65            &  8.19            &  8.56            &  ---             &  ---             \\
log R$_{23}$\footnote{For all the objects showing log $R_{23}$ $\ge$ 0.90, 12+log(O/H) = 8.20 is adopted in the $R23$ method (see section 3.2.2 for details)}  
  &  0.64             &  ---             &  1.00              &  0.92            &  0.96            &  0.26            &  0.84            &  0.56            &  0.93            &  0.90            \\
12+log(O/H)$_{R23 upper}$\footnote{O/H derived from the theoretical abundance calibration of $R_{23}$ according to McGaugh (1991): upper branch as parametrized in Kobulnicky et al. (\cite{k99}); values in parantheses were rejected. } 
 &  8.68             &  ---             &  8.20              &  8.20            &  8.20            &  8.98            &  (8.47)           &  8.78            &  8.20            &  8.20            \\
12+log(O/H)$_{R23 lower}$\footnote{O/H derived from the theoretical abundance calibration of $R_{23}$ according to McGaugh (1991): lower branch as parametrized in Kobulnicky et al. (\cite{k99}); values in parantheses were rejected. } 
 &  (8.15)           &  ---             &  8.20              &  8.20            &  8.20            &  (7.53)          &  8.28            &  (7.93)           &  8.20            &  8.20            \\
12+log(O/H)$_{ONS}$\footnote{O/H derived from Pilyugin, Vilchez and Thuan (2010) - ONS calibration} & 8.52 & --- & 7.81 & 8.09 & --- & 8.59 & 7.99 & 8.50 & --- & --- \\
12+log(O/H)$_{adopted}$ \footnote{The adopted O/H corresponds to the mean value obtained from the different derivations (see Section 4.2 for details)}
  &  8.60   &  8.55   &   8.00  &  8.16    & 8.20    &   8.78    &   8.17   &  8.63    & 8.20      & 8.20              \\
\\
\hline       
\end{tabular}
\end{center}
%\end{table}
%\end{tiny}
\end{sidewaystable}

%______________________________________________________________

\clearpage

\begin{table}[!t]
\begin{center}
\caption{Physical parameters of the galaxies observed in Hydra by Duc et al. (\cite{duc99,duc01}): (1) galaxy name, 
(2) total apparent magnitude in $K^\prime$, (3) central surface brightness in $K^\prime$ (mag/arcsec$^2$), 
(4) total absolute magnitude in $K^\prime$, (5) hydrogen velocity linewidth at $20\%$ height, $W_{20}$ (km/s), 
(6) oxygen metallicity (dex); (7) logarithm of galaxy gas mass, log M$_G$  (M$_\odot$); 
(8) logarithm of baryonic mass, log M$_B$ (M$_\odot$); (9) logarithm of the logarithm of the inverse gas fraction $\mu$. } 
\begin{tabular}{lrrrrrrrr}
\hline
\hline
\noalign{\smallskip}
$Galaxy$ & $m_{TK}$ & $\mu_{OK}$ & $M_{TK}$ & $W_{20}$ & 12+log(O/H) & log M$_G$ & log M$_{bary}$ & log log(1/$\mu$) \\
\hline
\noalign{\smallskip}

 H1033-2707 &  15.02  &  18.56  &  -18.31  &  116  &   8.01  &   8.93  &   9.08  &  -0.82 \\
 H1034-2758 &  14.61  &  19.00  &  -18.72  &  136  &   7.90  &   8.83  &   9.07  &  -0.61 \\
 H1038-2733 &  15.73  &  20.64  &  -17.60  &   67  &   8.04  &   8.87  &   8.96  &  -1.01 \\
 H1032-2638 &  15.62  &  20.23  &  -17.71  &  128  &   7.82  &    ---  &    ---  &   ---  \\

\noalign{\smallskip}
\hline
\hline
\end{tabular}
\label{table4}
\end{center}
\end{table}

%______________________________________________________________

% Figures

\clearpage

\clearpage

\begin{figure}[p]
\centering
\includegraphics[angle=0,width=14cm]{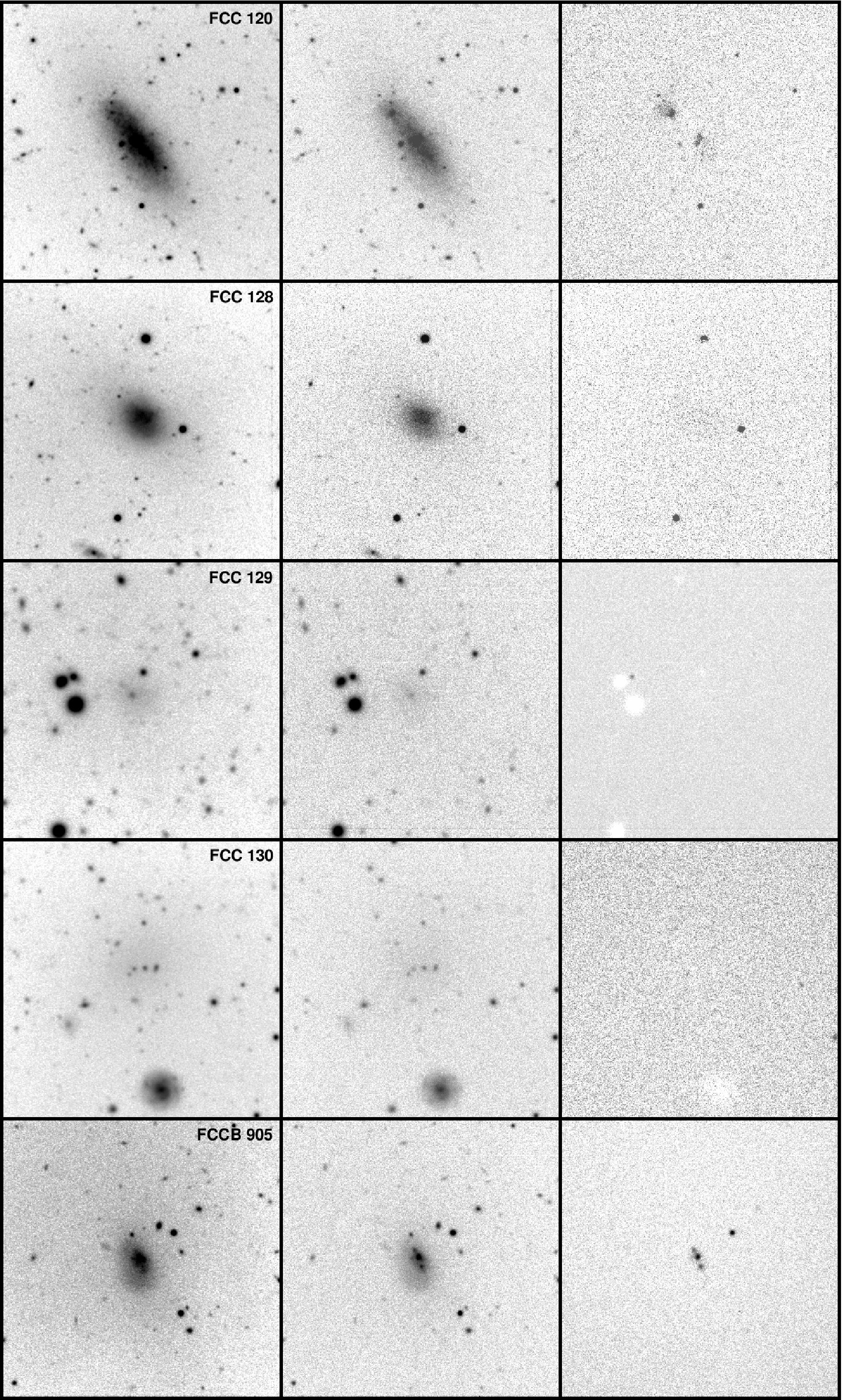}
\begin{center}
Figure 1 (continued - Fornax galaxies)
\end{center}
\end{figure}
\clearpage

\begin{figure}[p]
\centering
\includegraphics[angle=0,width=14cm]{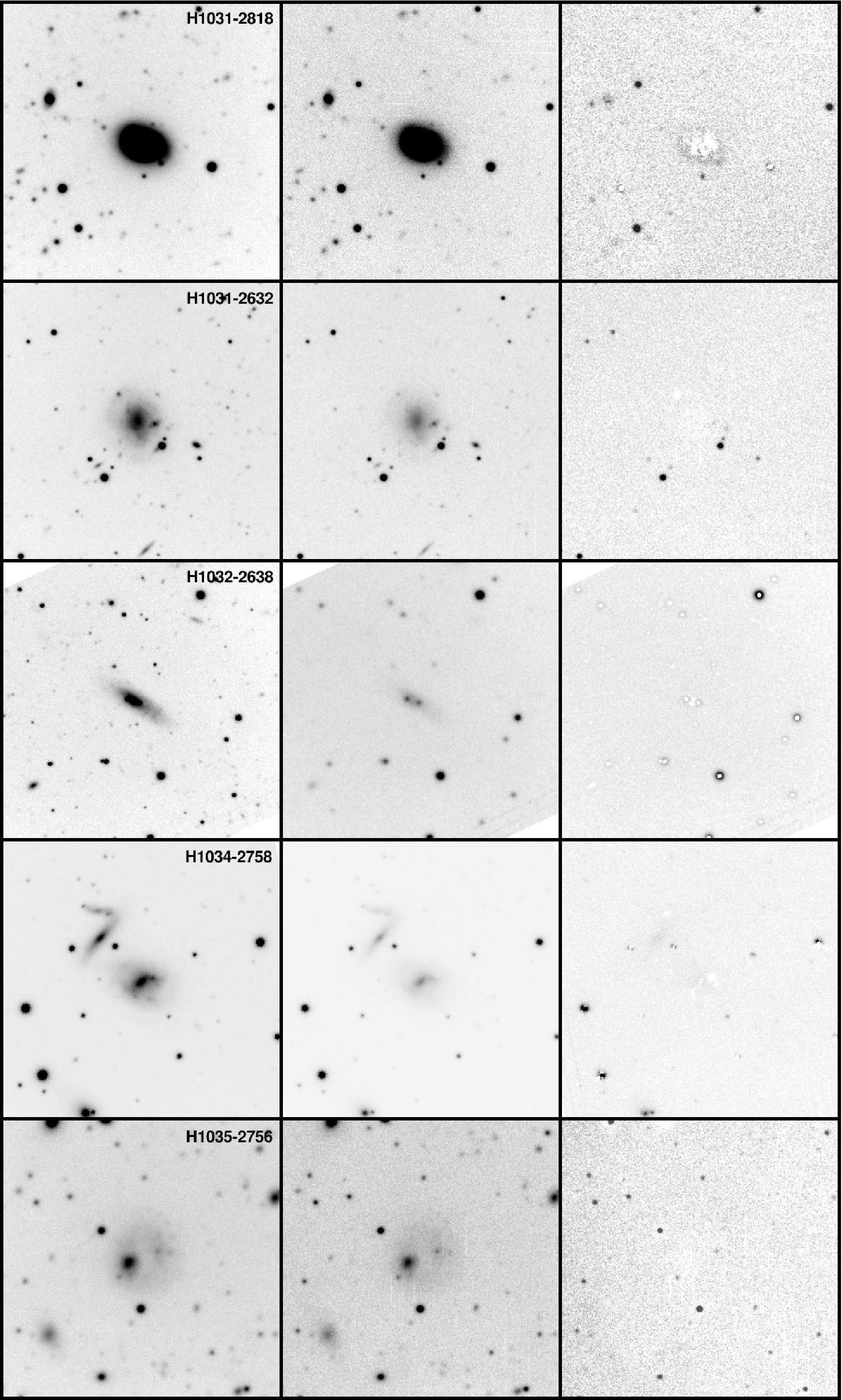}
\begin{center}
Figure 1 (continued - Hydra galaxies)
\end{center}
\end{figure}
\clearpage

\begin{figure}[p]
\centering
\includegraphics[angle=0,width=14cm]{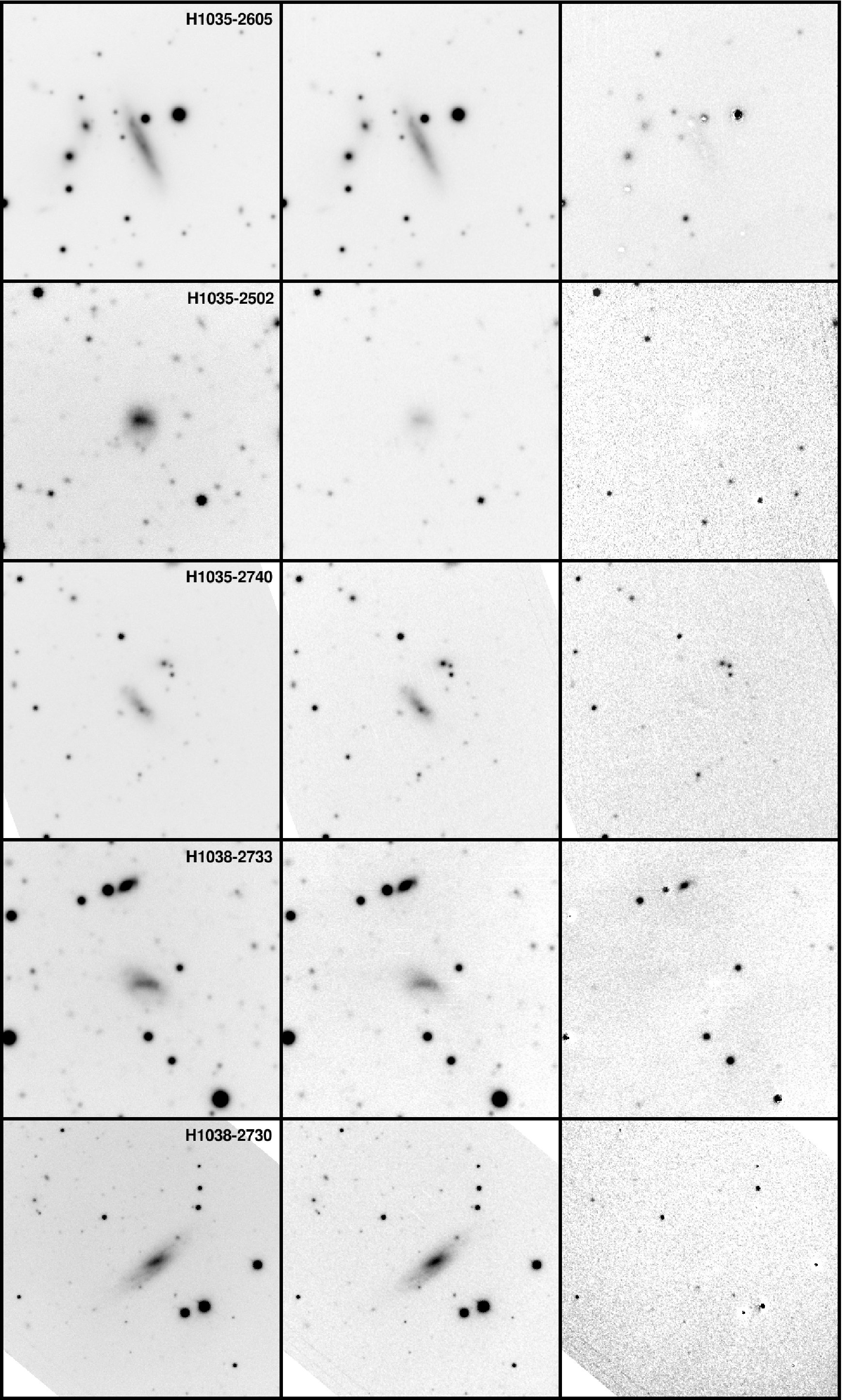}
\begin{center}
Figure 1 (continued - Hydra galaxies)
\end{center}
\end{figure}
\clearpage

\begin{figure}[p]
\centering
\includegraphics[angle=0,width=14cm]{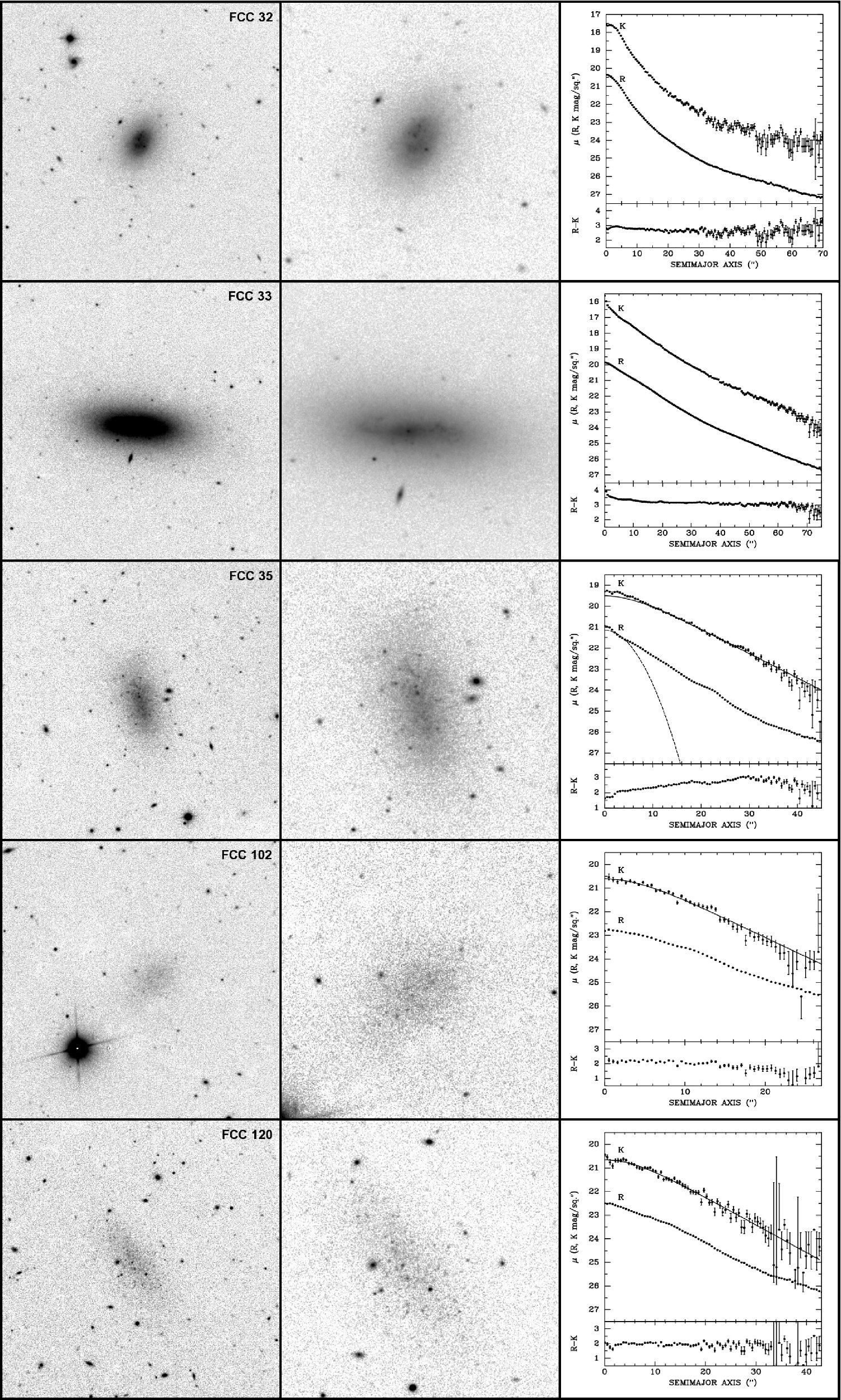}
\begin{center}
\caption{$K^\prime$ images and surface brightness profiles of BCD candidates observed at VLT. 
Left: $K^\prime$ image field of view $2\arcmin \times 2\arcmin$. 
Middle: Zoomed $K^\prime$ image showing the central core $1\arcmin \times 1\arcmin$ displayed in log scale. 
Right: Surface brightness profiles in $K^\prime$ (VLT imaging) and $R$ (Gemini pre-imaging), together with  
sech fits (continous line) and Gaussian fits (dashed line) of $K^\prime$ profiles (whenever available) and 
color profiles $R-K^\prime$ at bottom. All images use normal sky orientation. } 
\label{fig2}
\end{center}
\end{figure}
\clearpage

\begin{figure}[p]
\centering
\includegraphics[angle=0,width=14cm]{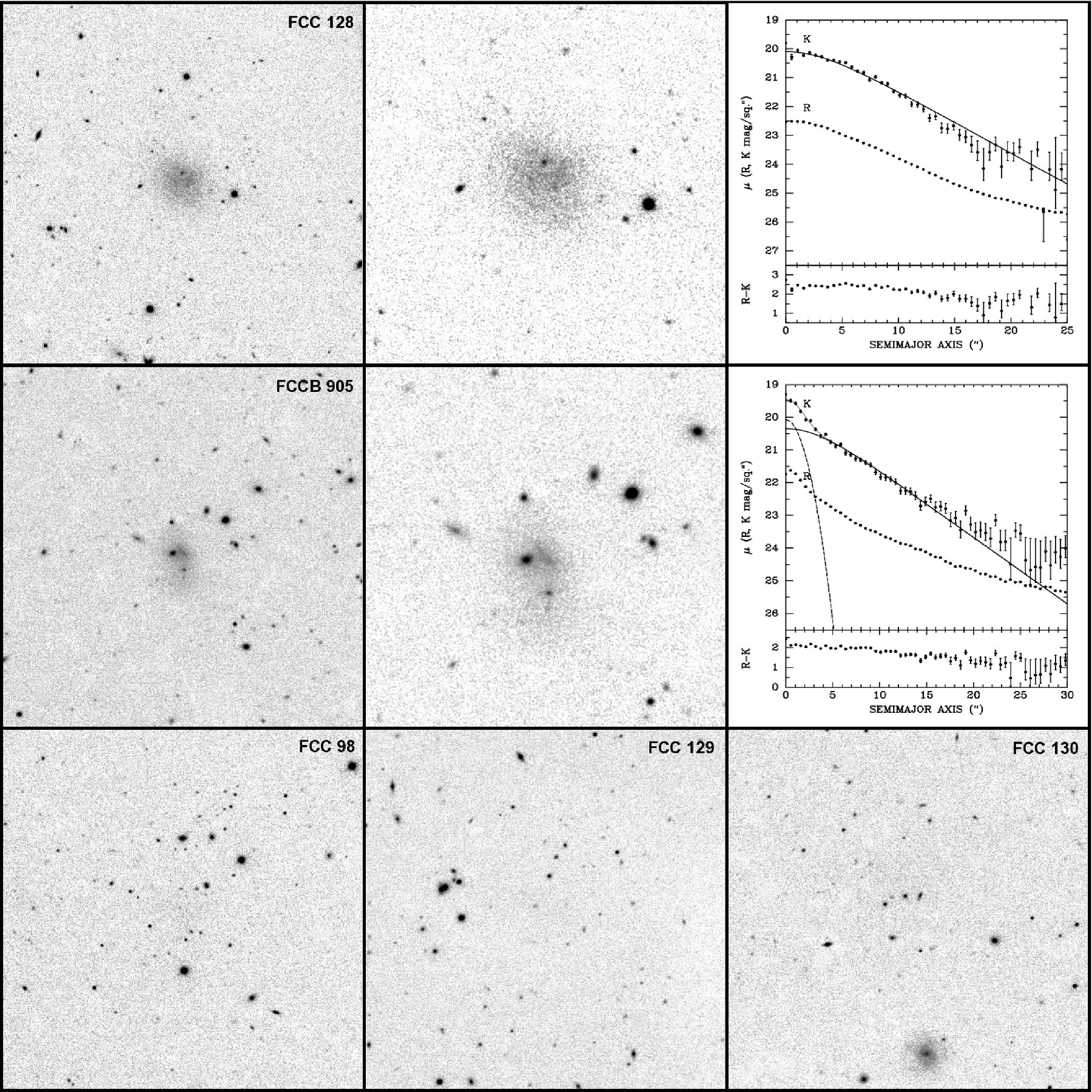}
\begin{center}
Figure 2 (continued - Fornax galaxies). The last three objects could not be detected. 
\end{center}
\end{figure}
\clearpage

\begin{figure}[p]
\centering
\includegraphics[angle=0,width=14cm]{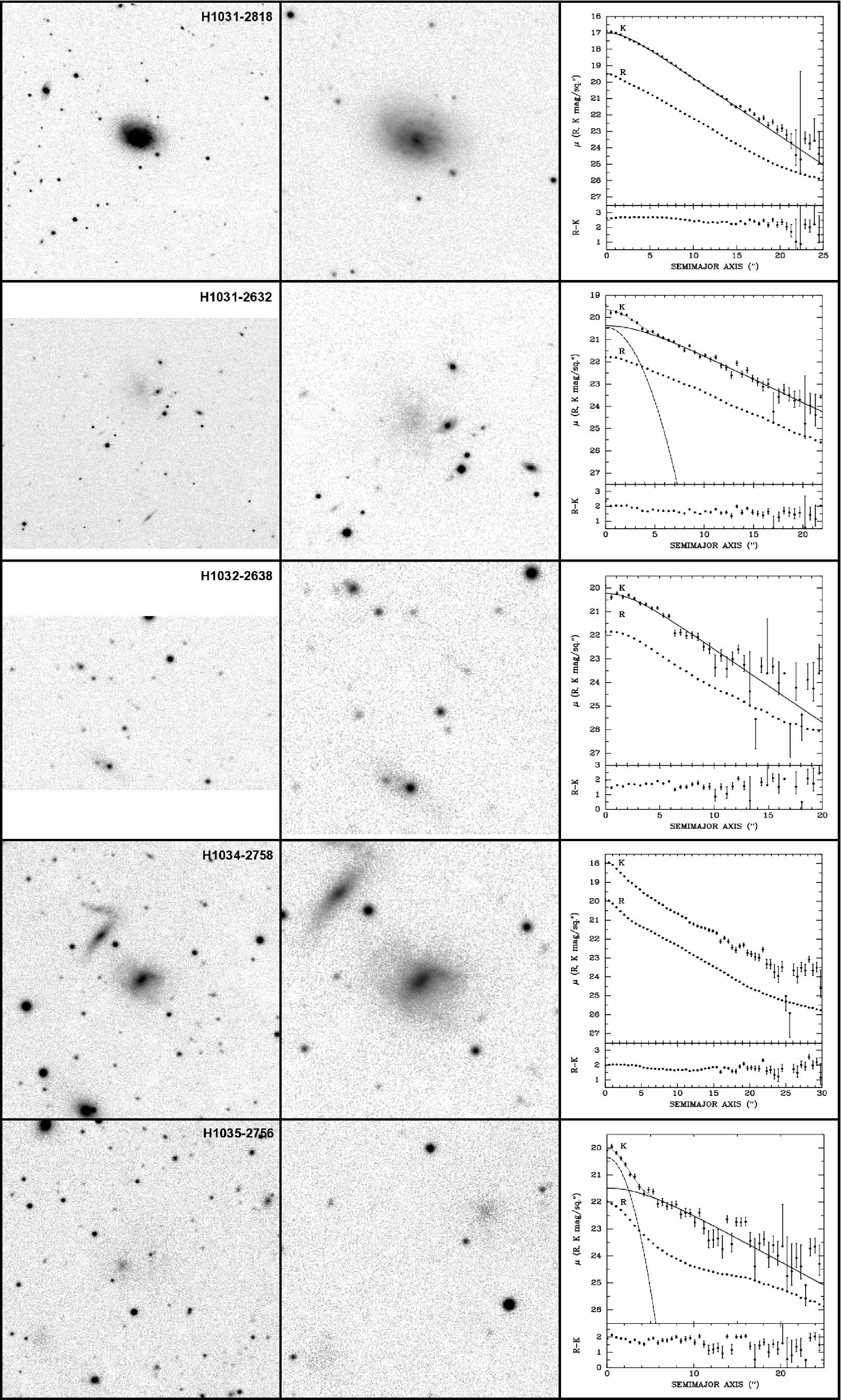}
\begin{center}
Figure 2 (continued - Hydra galaxies)
\end{center}
\end{figure}
\clearpage

\begin{figure}[p]
\centering
\includegraphics[angle=0,width=14cm]{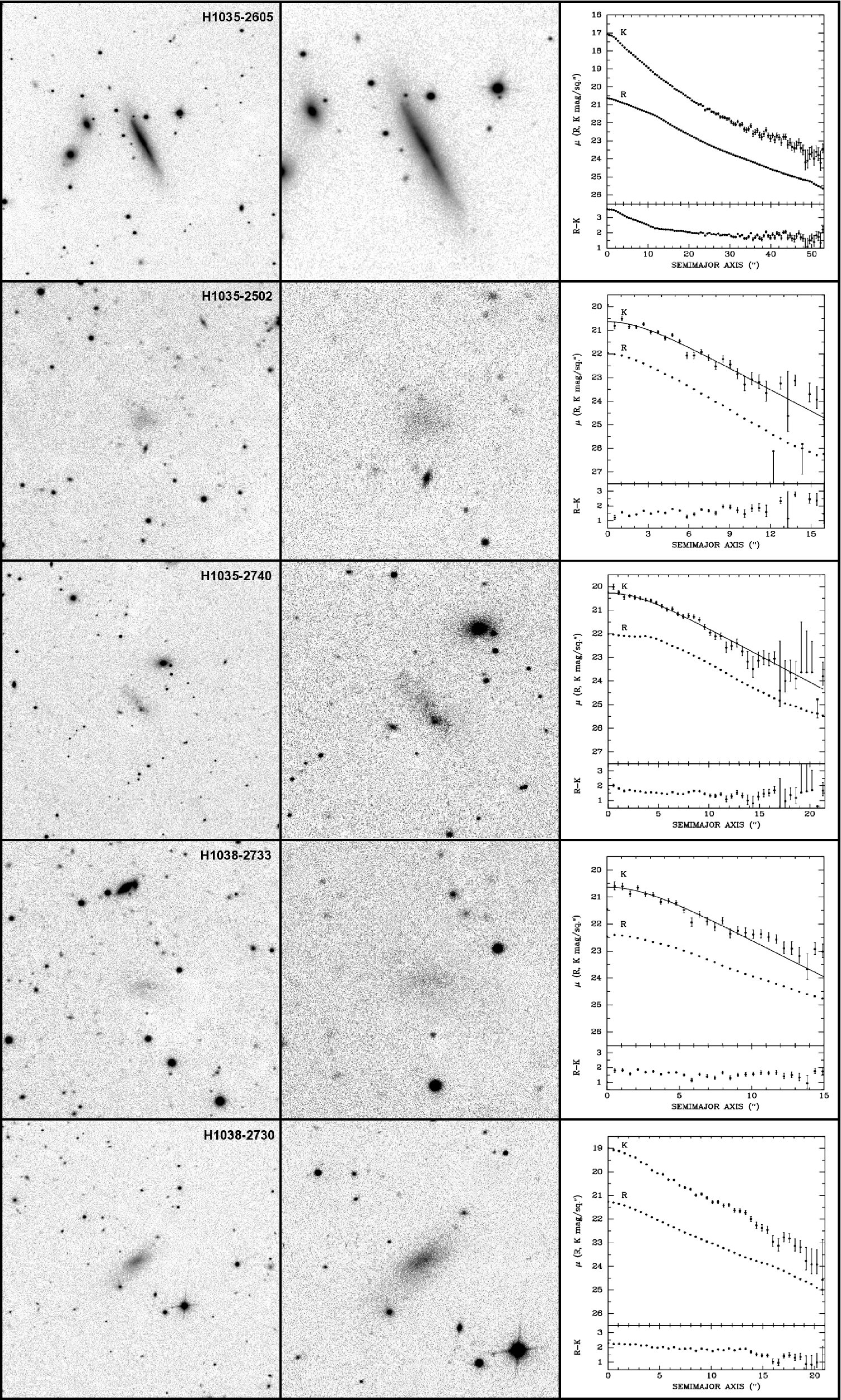}
\begin{center}
Figure 2 (continued - Hydra galaxies)
\end{center}
\end{figure}
\clearpage

\begin{figure}[p]
\centering
\includegraphics[angle=0,width=8cm]{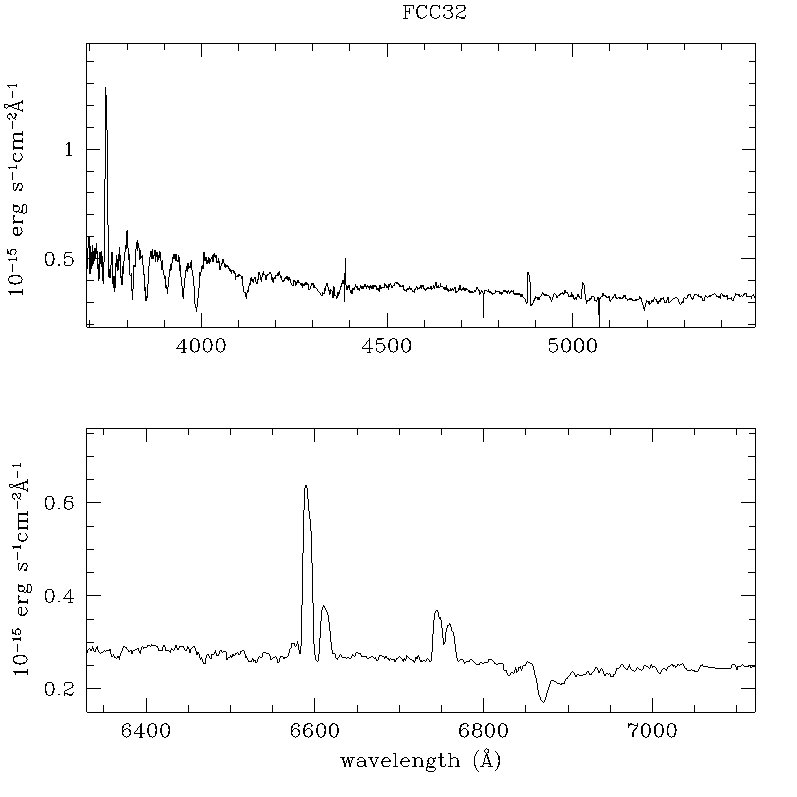}\includegraphics[angle=0,width=8cm]{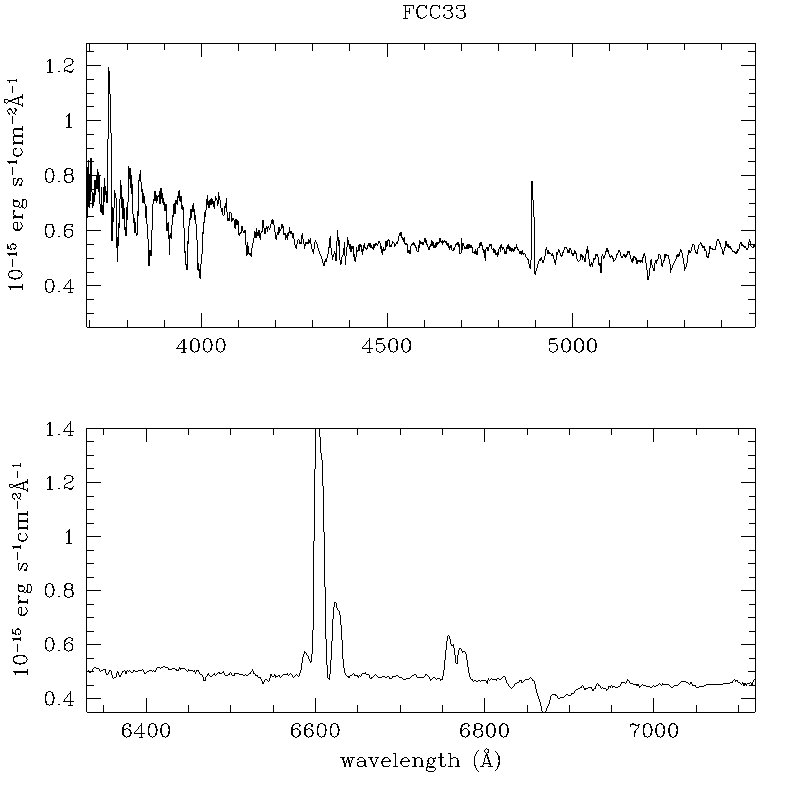}
\includegraphics[angle=0,width=8cm]{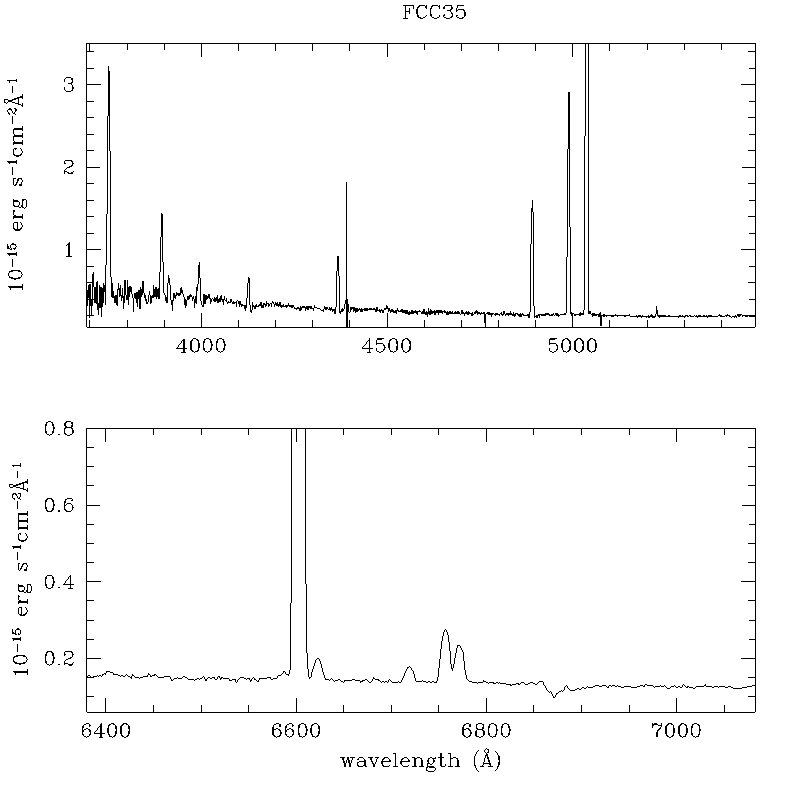}\includegraphics[angle=0,width=8cm]{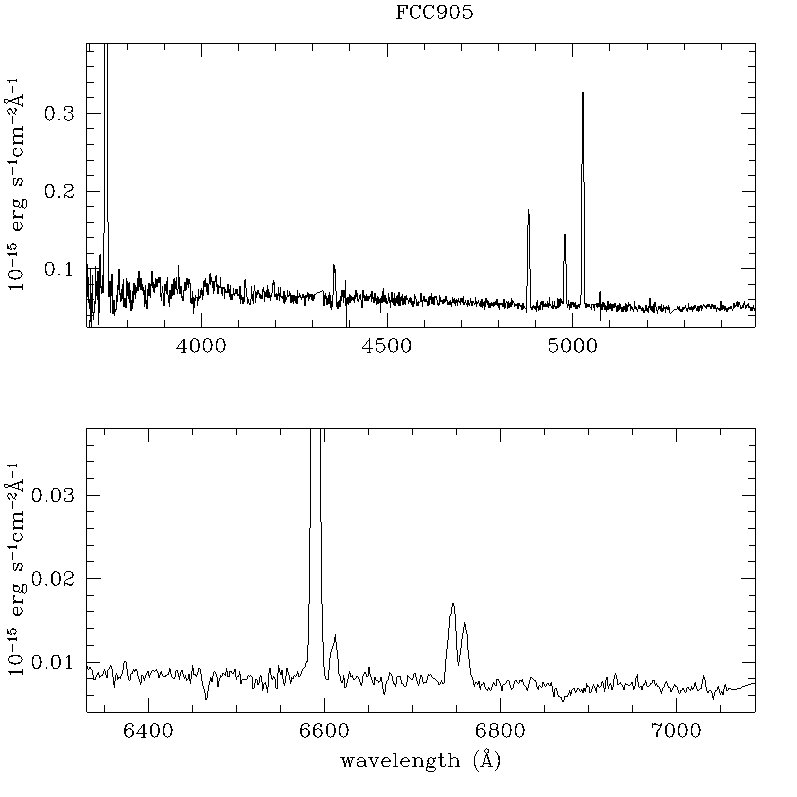}
\includegraphics[angle=0,width=8cm]{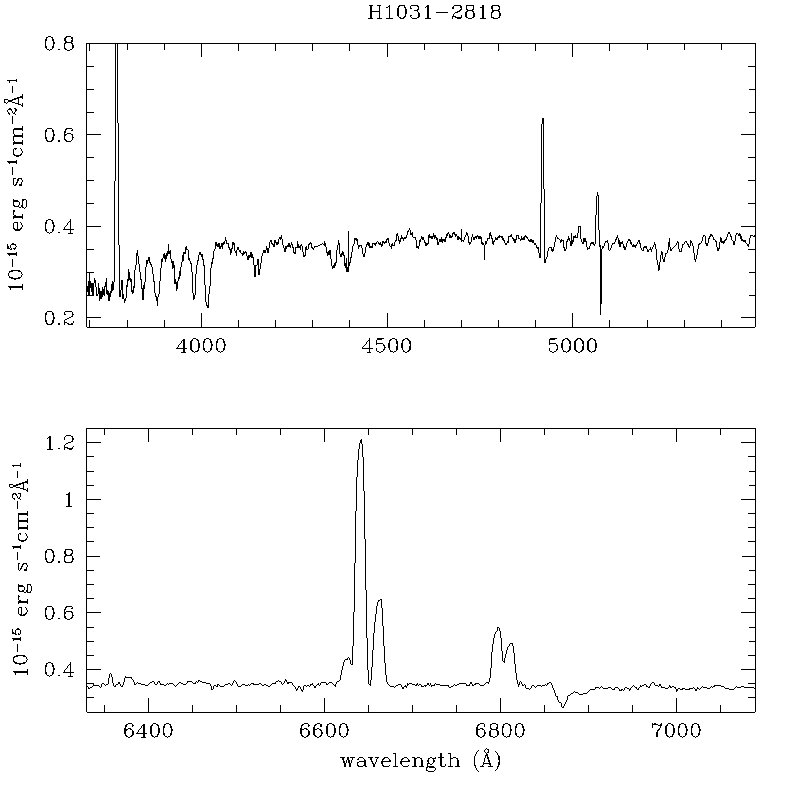}\includegraphics[angle=0,width=8cm]{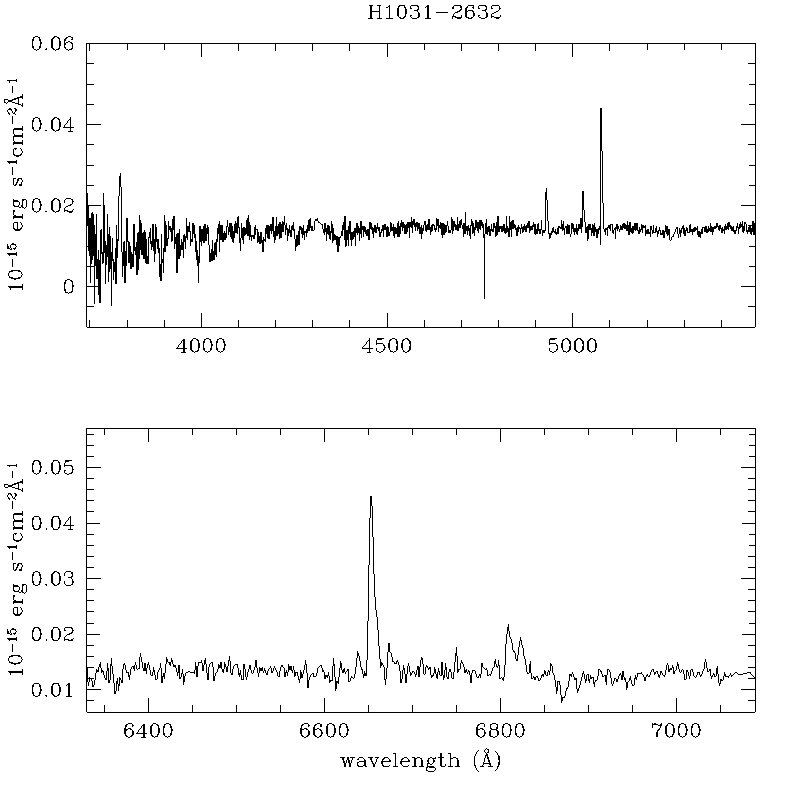}
\begin{center}
\caption{ 
Blue and red spectra (shown above and below), in units of 10$^{-15}$ erg s$^{-1}$ cm$^{-2}$ \AA$^{-1}$, 
of the star-forming candidates in our Hydra and Fornax samples. H1035-2756 and H1035-2740 were observed only in the blue side.}
\label{fig3}
\end{center}
\end{figure}
\clearpage

\begin{figure}[p]
\centering
\includegraphics[angle=0,width=8cm]{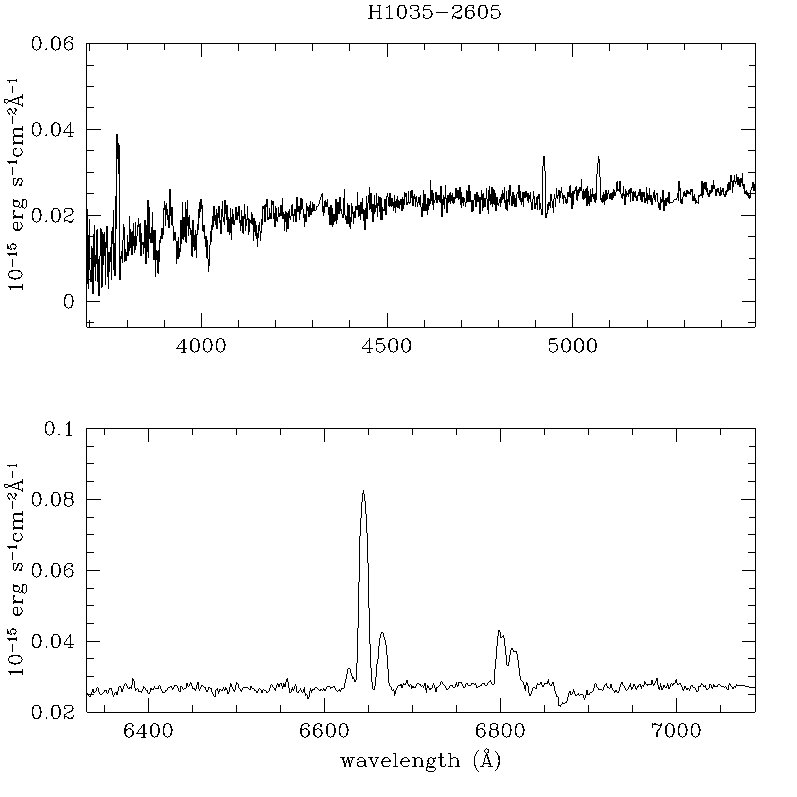}\includegraphics[angle=0,width=8cm]{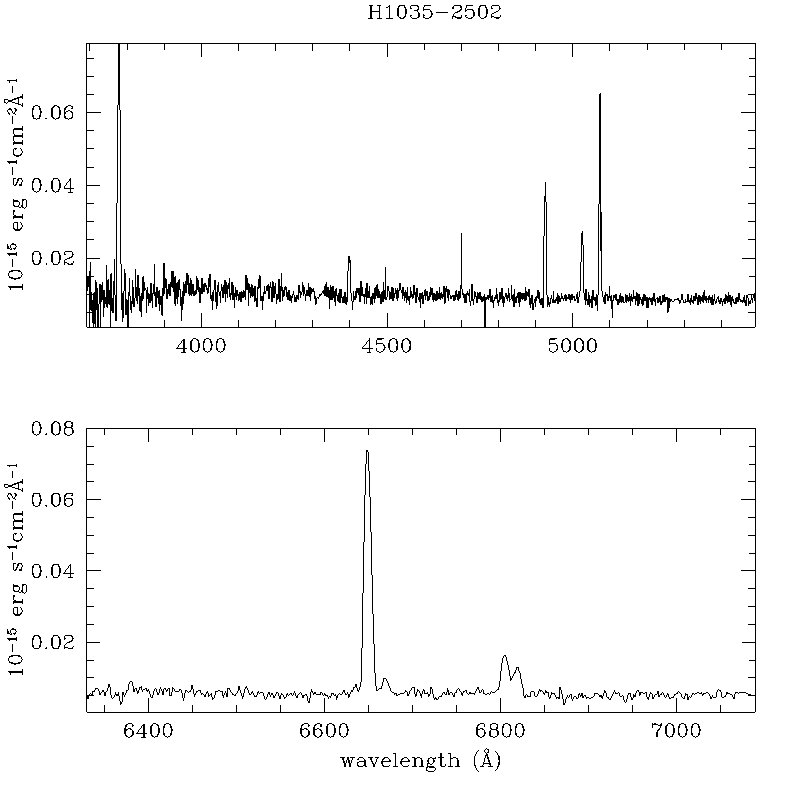}
\includegraphics[angle=0,width=8cm]{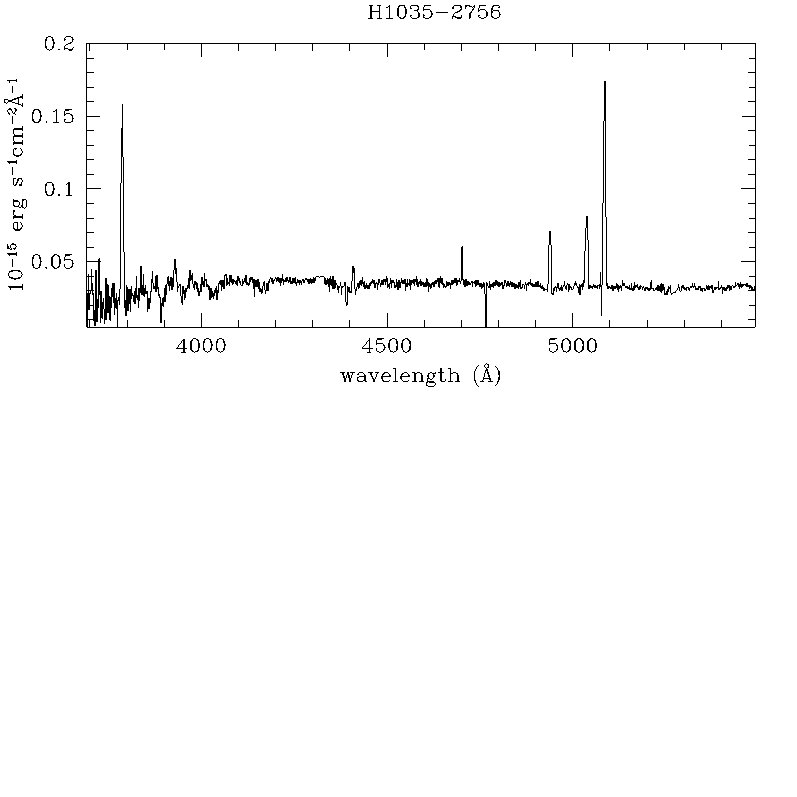}\includegraphics[angle=0,width=8cm]{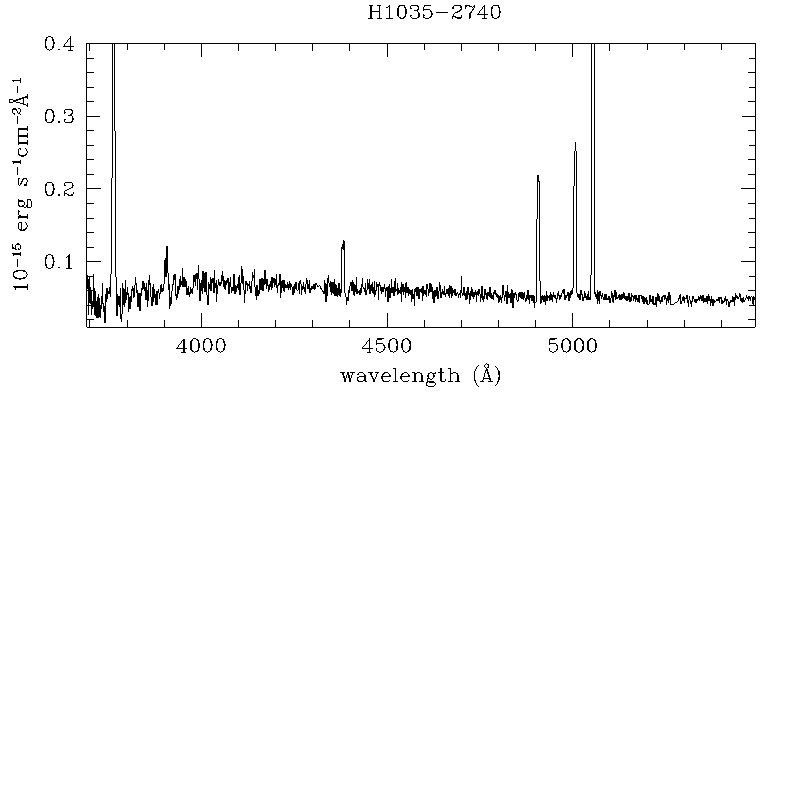}
\begin{center}
Figure 3 (continued) 
\end{center}
\end{figure}
\clearpage

\begin{figure}
\centering
\includegraphics[angle=0,width=9.5cm]{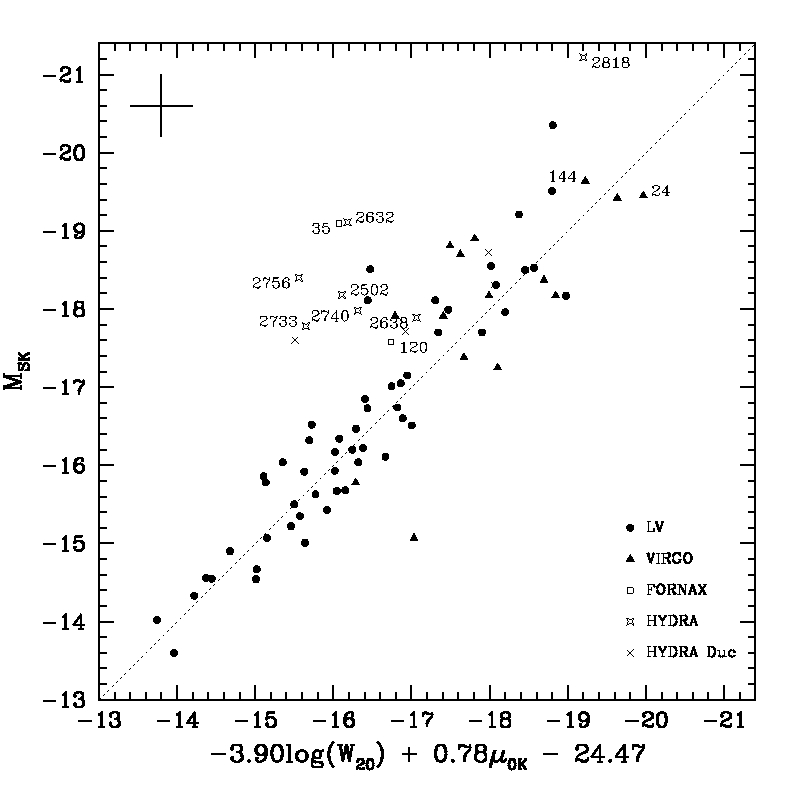}
\begin{center}
\caption{
The position of Fornax and Hydra galaxies with respect to the dwarf fundamental plane (FP) defined by 
50 star-forming dwarf galaxies from the LV on which Virgo star-forming dwarfs are also plotted 
(McCall et al. \cite{mcc11}). Last digits for Fornax and Hydra galaxies are labeled. Typical error bars 
of 0.4 mag are plotted in the upper-left corner. Only two objects appear close to the plane (one from 
each cluster), while most Hydra galaxies are located outside, suggesting some enviromental effects. 
} 
\label{fig4}
\end{center}
\end{figure}

\begin{figure}
\centering
\includegraphics[angle=0,width=9.5cm]{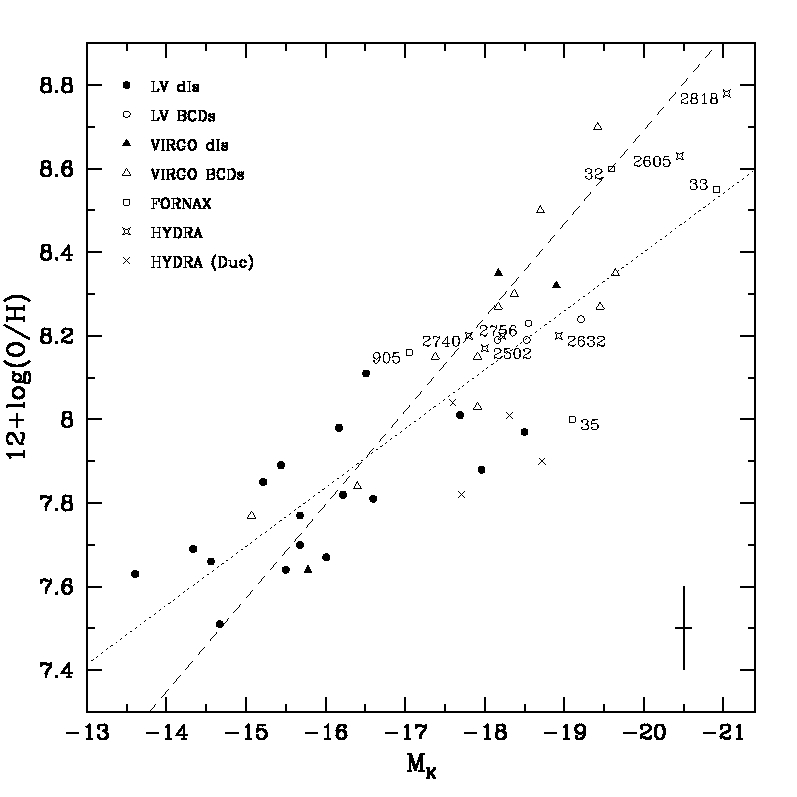}
\begin{center}
\caption{
The position of Fornax and Hydra galaxies on the luminosity-metallicity ($L-Z$) diagram populated by the 
LV star-forming dwarfs and Virgo star-forming dwarfs. The dotted line represents the fit obtained 
by Vaduvescu, McCall\& Richer (\cite{vad07}) using 25 LV and Virgo dIs, and the dashed line shows the fit 
obtained using 14 BCDs in LV and Virgo. Typical error bars of 0.1 dex and 0.1 mag are plotted. Most Fornax 
and Hydra objects appear confined to the $L-Z$ fits. } 
\label{fig5}
\end{center}
\end{figure}

\begin{figure}
\centering
\includegraphics[angle=0,width=9.5cm]{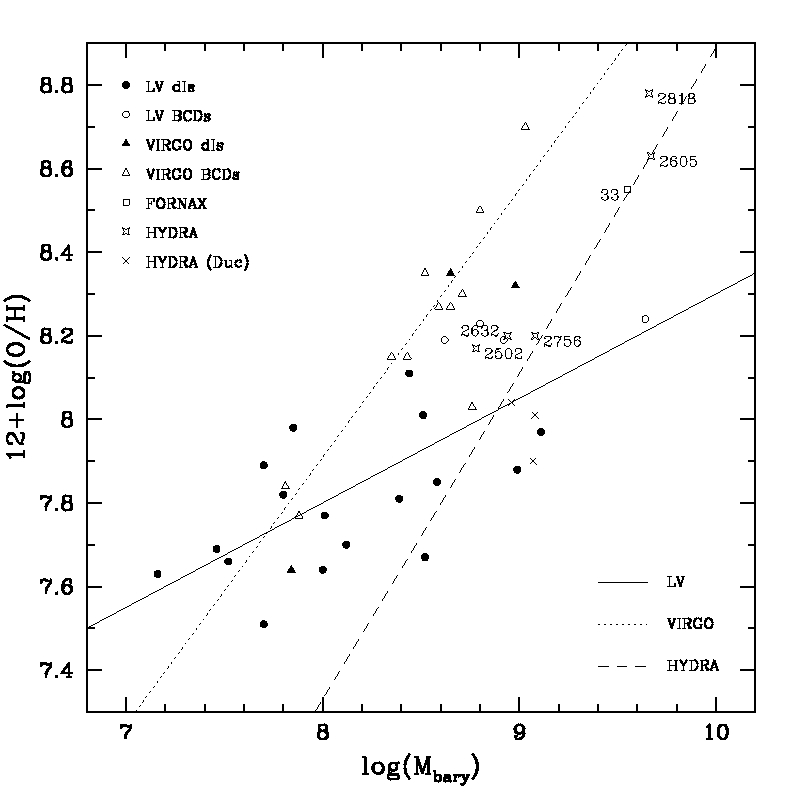}
\begin{center}
\caption{
The location of Fornax and Hydra dwarfs on the mass-metallicity diagram populated by the LV
star-forming dwarfs and Virgo star-forming dwarfs. With a solid line, we draw the LV fit for the 
star-forming galaxies (dIs and BCDs), with a dotted line the similar Virgo fit and with a dashed line 
the Hydra fit. Both Virgo and Hydra samples give a fit with a steeper slope than the LV one, suggesting 
that more metal-rich star-forming dwarfs tend to occupy regions of higher galaxy density. } 
\label{fig6}
\end{center}
\end{figure}

\begin{figure}
\centering
\includegraphics[angle=0,width=9.5cm]{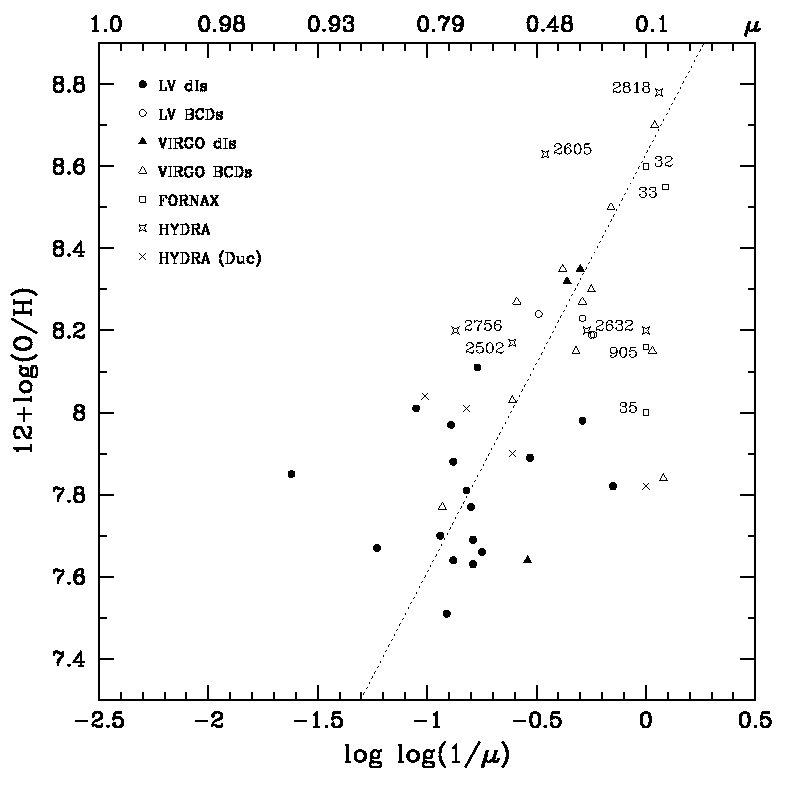}
\begin{center}
\caption{
The location of Fornax and Hydra dwarfs on the gas fraction-metallicity diagram populated by the Local 
Volume star-forming dwarfs and Virgo star-forming dwarfs. The dotted line represents the fit of Lee et al., 
\cite{lee03} consistent with the closed box model. Most Fornax and Hydra objects appear confined to this 
relation. } 
\label{fig7}
\end{center}
\end{figure}

\end{document}